\documentclass[manuscript,screen]{acmart}
\AtBeginDocument{%
  \providecommand\BibTeX{{%
    \normalfont B\kern-0.5em{\scshape i\kern-0.25em b}\kern-0.8em\TeX}}}

\setcopyright{acmcopyright}
\copyrightyear{2018}
\acmYear{2018}
\acmDOI{XXXXXXX.XXXXXXX}

\acmPrice{15.00}
\acmISBN{978-1-4503-XXXX-X/18/06}

\usepackage{amsmath}

\usepackage{balance}  
\usepackage{graphics} 
\usepackage{times}    
\usepackage{color}
\usepackage{textcomp}
\usepackage{booktabs}
\usepackage{ccicons}

\usepackage{url}
\usepackage{fancybox}
\usepackage{multirow}
\usepackage{flushend}
\usepackage{booktabs}
\usepackage{tabularx}
\usepackage{comment}
\usepackage{array}
\usepackage[flushleft]{threeparttable}
\usepackage{mdframed}
\graphicspath{{}{images/}{dia/}}
\DeclareGraphicsExtensions{.pdf,.png}
\usepackage{subfig}

\usepackage{listings}
\usepackage{courier}
\usepackage{hyperref}

\usepackage{xpatch}

\xpatchcmd{\refstepcounter}{%
  \stepcounter{#1}%
}{%
  \stepcounter{#1}%
  \xdef\scr{\number\value{#1}}%
}{\typeout{success}}{\typeout{failure}}

\newcounter{o}
\usepackage{soul}
\usepackage{tikz}
\definecolor{1c1}{RGB}{188,162,6}
\definecolor{1c2}{RGB}{137,129,80}
\definecolor{1c3}{RGB}{239,167,31}
\definecolor{1c4}{RGB}{88,194,241}
\definecolor{1c5}{RGB}{6,180,188}

\tikzset{mynode/.style={draw=white,solid,circle,fill=green,inner sep=1pt, thick,
text=black}}
\tikzset{arrow line/.style={dashed, line width= 2.5pt, color=#1}}

\def\bf{\textbf}

\def\sec {Section~}

\def\it{\textit}

\def\tt{\mct}

\newcommand{\mct}[1]{{\footnotesize {\texttt {#1}}}}

\usepackage{paralist}

\newcommand{\nd}{\vspace{1mm}\noindent}
\usepackage{tikz}

\lstdefinestyle{inlinecode}{basicstyle={\ttfamily\scriptsize\bfseries}}

\newcommand{\urls}[1]{{\scriptsize\url{#1}}}
\usepackage{tcolorbox}

\newcommand{\rev}[1]{\textcolor{black}{#1}}

\newcommand{\dq}[1]{\href{https://stackoverflow.com/questions/#1/}{$Q_{#1}$}}
\newcommand{\da}[1]{\href{https://stackoverflow.com/a/#1/}{$A_{#1}$}}

\usepackage{paralist}
\usepackage[outercaption]{sidecap}
\usepackage [autostyle, english = american]{csquotes}
\MakeOuterQuote{"}
\newcounter{scn}
\usepackage[shortlabels]{enumitem}
\usepackage{tikz}
\usepackage{styles/pgf-pie}
\usetikzlibrary{positioning,shadows}
\usepackage{balance}

\newif\ifpienumberinlegend
\pgfkeys{/number in legend/.code=
    \expandafter\let\expandafter\ifpienumberinlegend
    \csname if#1\endcsname
    \ifpienumberinlegend

    \def\beforenumber##1\afternumber{}%
    \fi,
    /number in legend/.default=true
}
\definecolor{1c1}{RGB}{188,162,6}
\definecolor{1c2}{RGB}{137,129,80}
\definecolor{1c3}{RGB}{239,167,31}
\definecolor{1c4}{RGB}{88,194,241}
\definecolor{1c5}{RGB}{6,180,188}

\tikzset{mynode/.style={draw=white,solid,circle,fill=green,inner sep=1pt, thick,
text=black}}
\tikzset{arrow line/.style={dashed, line width= 2.5pt, color=#1}}

\usepackage{bchart}

\usepackage{xcolor}
\definecolor{ao(english)}{rgb}{0.0, 0.5, 0.0}

\def\test#1{%
\ifnum0#1>0
      #1
    \fi
}

\newcommand{\sixbars}[6]{
{{\color{black}\rule{#1pt}{4pt}} \test{#1}}
{{\color{ao(english)}\rule{#2pt}{4pt}} \test{#2}}
{{\color{magenta}\rule{#3pt}{4pt}} \test{#3}}
{{\color{red}\rule{#4pt}{4pt}} \test{#4}}
{{\color{cyan}\rule{#5pt}{4pt}} \test{#5}}
{{\color{orange}\rule{#6pt}{4pt}} \test{#6}}
}

\newcommand{\twobars}[2]{
{{\color{black}\rule{#1pt}{4pt}} \test{#1}}
{{\color{ao(english)}\rule{#2pt}{4pt}} \test{#2}}
}

\begin{document}
\title[Practitioners' Discussions of the AutoML Tool/Platforms]{Challenges and Barriers of Using Low Code Software for Machine Learning}


\author{Md Abdullah Al Alamin}
\email{mdabdullahal.alamin@ucalgary.ca}
\author{Gias Uddin}
\email{gias.uddin@ucalgary.ca}
\affiliation{%
  \institution{DISA Lab, University of Calgary}
  \streetaddress{2500 University Dr NW}
  \city{Calgary}
  \state{Alberta}
  \country{Canada}
  \postcode{T2N 1N4}
}

\begin{abstract}
As big data grows ubiquitous across many domains, more and more stakeholders seek to develop Machine Learning (ML) applications on their data. The success of an ML application usually depends on the close collaboration of ML experts and domain experts. However, the shortage of ML engineers remains a fundamental problem. Low-code Machine learning tools/platforms (aka, AutoML) aim to democratize ML development to domain experts by automating many repetitive tasks in the ML pipeline, such as data collection, data pre-processing, feature engineering, model design, optimal hyper-parameter configuration, and model evaluation.
However, even with minimal the hand coding support via the end-to-end pipeline of AutoML tools, it still requires human involvement in vital steps such as understanding the problem scope, domain-specific data, designing an appropriate training and testing dataset, model deployment \& monitoring.
As AutoML has some unique characteristics to traditional ML, it is vital to study the challenges ML practitioners face while they use the currently available AutoML tools, so that appropriate measures can be taken to address the challenges and to realize the vision of democratizing ML using low code software development principles and methodologies.

This research presents an empirical study of around 14k posts (questions + accepted answers) from Stack Overflow (SO) that contained AutoML-related discussions. Software developers frequently use the online developer Q\&A site SO to seek technical assistance. We observe a growing number of AutoML-related discussions in SO. We use LDA Topic Modeling to determine the topics discussed in those posts. Additionally, we examine how these topics are spread across the various Machine Learning Life Cycle (MLLC) phases and their popularity and difficulty. This study offers several interesting findings. First, we find 13 AutoML topics that we group into four categories. The MLOps topic category (43\% questions) is the largest, followed by Model (28\% questions), Data (27\% questions), Documentation (2\% questions).
Second, Most questions are asked during Model training (29\%) (i.e., implementation phase) and Data preparation (25\%) MLLC phase.
Third, AutoML practitioners find the MLOps topic category most challenging, especially topics related to model deployment \& monitoring and Automated ML pipeline.
Fourth, the Requirement analysis and scope definition MLLC phase are the most popular and challenging for AutoML practitioners. They also find the Model deployment and Model evaluation MLLC phase more complex than other phases like Data preparation and Model training.
Fifth, MLOps topic category and Model deployment and Monitoring phase are more predominant and popular in cloud-based AutoML solution and Model topic category whereas Model evaluation phase is more dominant and popular in non-cloud AutoML solutions.
These findings have implications for all three AutoML stakeholders: AutoML researchers, AutoML service vendors, and AutoML developers. Academia and Industry collaboration can improve different aspects of AutoML, such as better DevOps/deployment support and tutorial-based documentation.

\end{abstract}




\maketitle

\section{Introduction}

With the rise of automation and digitization, the term ``big data'' has become ubiquitous across a wide range of industries. The rapid growth of cloud computing and tremendous progress in machine learning has attracted businesses to hire ML engineers to make the most value out of their data. ML has grown in popularity as a solution to various software engineering issues, including speech recognition, image processing, natural language processing etc. Nevertheless, The development of an ML model requires significant human expertise. It is necessary to use intuition based on prior knowledge and experience to determine the best ML algorithm/architecture for each dataset. This high demand of ML expertise coupled with the shortage of ML developers and the repetitive nature of many ML pipeline processes have inspired the adoption of low-code machine learning (i.e., AutoML) tools/platforms.

The primary goal of low-code AutoML services is to reduce the manual efforts of developing different ML pipelines, which thus helps in accelerating their development and deployment. 
This is necessary due to the fact that a business may have a specific business requirement and the domain experts/stakeholders in the business know it, but they lack the skills to develop an ML application. Indeed, for ML development, there are often primarily two key users as follows.
\begin{enumerate}
    \item \bf{Domain Experts} are knowledgeable about the problem domain (e.g., rain forecasting, cancer diagnosis, etc.) where the ML is applied. So that they have a deep understanding of the scope of the problem and the dataset.
    \item \bf{ML Experts} are experienced with the nuances of the ML model. They have a greater understanding and experience in selecting the appropriate ML algorithm/architecture, engineering features, training the model, and evaluating its performance.
\end{enumerate}
Both of the domain and ML experts are equally important for the success of the ML-based solution development, and are often complementary to each other during the design and development of the solutions.
AutoML aims to make ML more accessible to domain experts by providing end-to-end abstraction from data filtering to model designing \& training to model deployment \& monitoring. AutoML is becoming an essential topic for software engineering researchers as more and more AutoML applications are developed and deployed. AutoML research is progressing rapidly as academia and industry work collaboratively by making ML research their priority. For example, currently, in at least some experimental settings, AutoML tools yield better results than manually designed models by ML experts~\cite{Mazzawi2019ImprovingKS}
 (c.f. background of AutoML in \sec\ref{sec:background}).

Machine learning practitioners face many challenges because of the interdisciplinary nature of the ML domain~\cite{alshangiti2019developing}. They require not only the software engineering expertise to configure and set up new libraries but also in data visualization, linear algebra and statistics.
There have been quite some studies~\cite{islam2019developers, sculley2015hidden} on the challenges of deep learning and ML tools, but there are no formal studies on the challenges AutoML practitioners have asked about on public forums. There are also studies on developers' discussion on machine learning domain~\cite{bangash2019developers, humbatova2020taxonomy, alshangiti2019developing} and deep learning frameworks~\cite{islam2019developers, han2020programmers, zhang2018empirical, cummaudo2020interpreting, chen2020comprehensive}, but \textit{there are no studies on low-code, i.e., AutoML tools/platforms (c.f. related work in \sec\ref{sec:related_work}).}

The online developer forum Stack Overflow (SO) is the most popular Q\&A site with around 120 million posts and 12 million registered users~\cite{website:stackoverflow}. Several research has conducted by analyzing developers' discussion of SO posts (e.g., traditional low-code practitioners discussion~\cite{alamin_LCSD_EMSE, alamin2021empirical}, IoT~\cite{iot21}, big data~\cite{bagherzadeh2019going}, blockchain~\cite{wan2019discussed}, docker developers challenges~\cite{haque2020challenges}, concurrency~\cite{ahmed2018concurrency}, microservices~\cite{bandeira2019we}) etc.
\textit{The increasing popularity and distinctive character of low-code ML software development approaches (i.e., AutoML tools/frameworks) make it imperative for the SE research community to explore and analyze what practitioners shared publicly.}

To that end, in this paper, we present an empirical study of 14.3K SO posts (e.g., 10.5k Q + 3.3k Acc Ans) relating to AutoML-related discussion in SO  to ascertain the interest and challenges of AutoML practitioners (c.f. study methodology in \sec\ref{sec:methodology}). We use SOTorrent dataset~\cite{SOdump} and Latent Dirichlet Allocation (LDA) to systematically analyze practitioners' discussed topics similar to other studies~\cite{alamin_LCSD_EMSE, iot21, alamin2021empirical, abdellatif2020challenges, ahmed2018concurrency}. We explore the following four research questions by analyzing the dataset (c.f. results in \sec\ref{sec:results}).

\nd\bf{RQ1. What topics do practitioners discuss regarding AutoML services on SO?}
As low-code machine learning (i.e., AutoML) is an emerging new paradigm, it is vital to study publicly shared queries by AutoML practitioners on a Q\&A platform such as SO. We extract 14.3K AutoML related SO posts and apply the LDA topic modelling method~\cite{blei2003latent} to our dataset. We find a total of 13 AutoML-related topics grouped into four categories: MLOps (43\% Questions, 5 topics),  Model (28\% Questions, 4 topics), Data (27\% Questions, 3 topics), and Documentation (2\% Questions, topic). We find that around 40\% of the questions are related to supported features of a specific AutoML solution providers, around 15\% of questions are related to model design and development, and 20\% of questions are related to data pre-processing and management. We find relatively fewer questions on programming, and details configuration on ML algorithm as AutoML aims to provide a higher level abstraction over data processing and model design and development pipeline. However, AutoML practitioners still struggle with development system configuration and model deployment.

\nd\bf{RQ2. How AutoML topics are distributed across machine learning life cycle phases?}
AutoML aims to provide an end-to-end pipeline service to streamline ML development. The successful adoption of this technology largely depends on how effective this is on different machine learning life cycle (MLLC) phases. So, following related studies~\cite{alshangiti2019developing, al2021quality, islam2019developers}, we manually analyze and annotate 348 AutoML questions into one of six ML life cycle stages by taking statically significant questions from all the four topic categories. We find that Model training, i.e., the implementation phase is the most dominant (28.7\% questions), followed by Data Preparation (24.7\% questions) and Model Designing (18.7\% questions).

\nd\bf{RQ3. What AutoML topics are most popular and difficult on SO?}
From our previous research questions, we find that AutoML practitioners discuss diverse topics in different stages of machine learning development. However, some of these questions are popular and have larger community support. We find that AutoML practitioners find the MLOps topic category most popular and challenging. AutoML practitioners find the Requirement Analysis phase the most challenging and popular, followed by the model deployment \& monitoring MLLC phase. We find that AutoML practitioners find Model Deployment \& Load topic to be most challenging regarding the percentage of questions without accepted answers and median hours required to get accepted answers. We also find that questions related to improving the performance of AutoML models are most popular among AutoML practitioners regarding average view count and average score.

\nd\bf{RQ4. How does the topic distribution between cloud and non-cloud AutoML services differ?}
Our analyzed dataset contains both cloud-based and non-cloud-based AutoML services. Cloud-based solutions provides an end-to-end pipeline for data collecting to model operationalization, whereas non-cloud-based services offer greater customizability for data preparation and model development. 
We find that, in our dataset around 65\% SO questions  belong to cloud and 35\% belong to non-cloud based AutoML solution.
Cloud-based AutoML solution predominate over Model Deployment and Monitoring phase (82\%) and non-cloud based AutoML solution predominate over Model Evaluation phase.
MLOps topic category (i.e., 80\%) is predominated over cloud-based AutoML solutions, while Model (i.e., 59\%) topic category is predominated by non-cloud based AutoML solutions.

Our study findings offer valuable insights to AutoML researchers, service providers, and educators regarding what aspect of AutoML requires improvement from the practitioners' perspective (c.f. discussions and implications in \sec\ref{sec:discussion}).
Specifically, our findings can enhance our understanding of the AutoML practitioners' struggle and help the researchers and platform vendors better focus on the specific challenges. For example, AutoML solutions lack better support for deployment, and practitioners can prepare for potentially challenging areas. In addition, all stakeholders and practitioners of AutoML can collaborate to provide enhanced documentation and tutorials. The AutoML service vendors can better support model deployment, monitoring and fine-tuning.


\nd\bf{Replication Package}: The code and data are shared in \url{https://github.com/disa-lab/automl-challenge-so}

\section{Background} \label{sec:background}

\subsection{AutoML as Low-code Tool/Platform for ML}
Recent advancements in machine learning (ML) have yielded highly promising results for a variety of tasks, including regression, classification, clustering, etc., on diverse dataset types (e.g., texts, images, structured/unstructured data). The development of an ML model requires significant human expertise. Finding the optimal ML algorithm/architecture for each dataset necessitates intuition based on past experience. ML-expert and domain-expert collaboration is required for these laborious and arduous tasks. The shortage of ML engineers and the tedious nature of experimentation with different configuration values sparked the idea of a low-code approach for ML. This low-code machine learning solution seeks to solve this issue by automating some ML pipeline processes and offering a higher level of abstraction over the complexities of ML hyperparameter tuning, allowing domain experts to design ML applications without extensive ML expertise. It significantly increases productivity for machine learning practitioners, researchers, and data scientists.
The primary goal of low-code AutoML tools is to reduce the manual efforts of different ML pipelines, thus accelerating their development and deployment.

In general terms, a machine learning program is a program that can learn from experience, i.e., data. In the traditional approach, a human expert analyses data and explores the search space to find the best model. AutoML~\cite{he2021automl, das2017survey} aims to democratize machine learning to domain experts by automating and abstracting machine learning-related complexities. It aims to solve the challenge of automating the Combined Algorithm Selection and Hyper-parameter tuning (CASH) problem. AutoML is a combination of automation and ML. It automates various tasks on the ML pipeline such as data prepossessing, model selection, hyper-parameter tuning, and model parameter optimization. They employ various types of techniques such as grid search, genetics, and Bayesian algorithms. Some AutoML services also help with data visualization, model interpretability, and deployment. It helps non-ML experts develop ML applications and provides opportunities for ML experts to engage in other tasks~\cite{gijsbers2019open}. The lack of ML experts and exponential growth in computational power make AutoML a hot topic for academia and the industry. Research on AutoML research is progressing rapidly; in some cases, at least in the experimental settings, AutoML tools are producing the best hand-designed models by ML experts~\cite{Mazzawi2019ImprovingKS}. 

\subsection{AutoML Approaches}\label{sub-sec:automl}

\begin{figure}[t]
\centering
\includegraphics[scale=0.55]{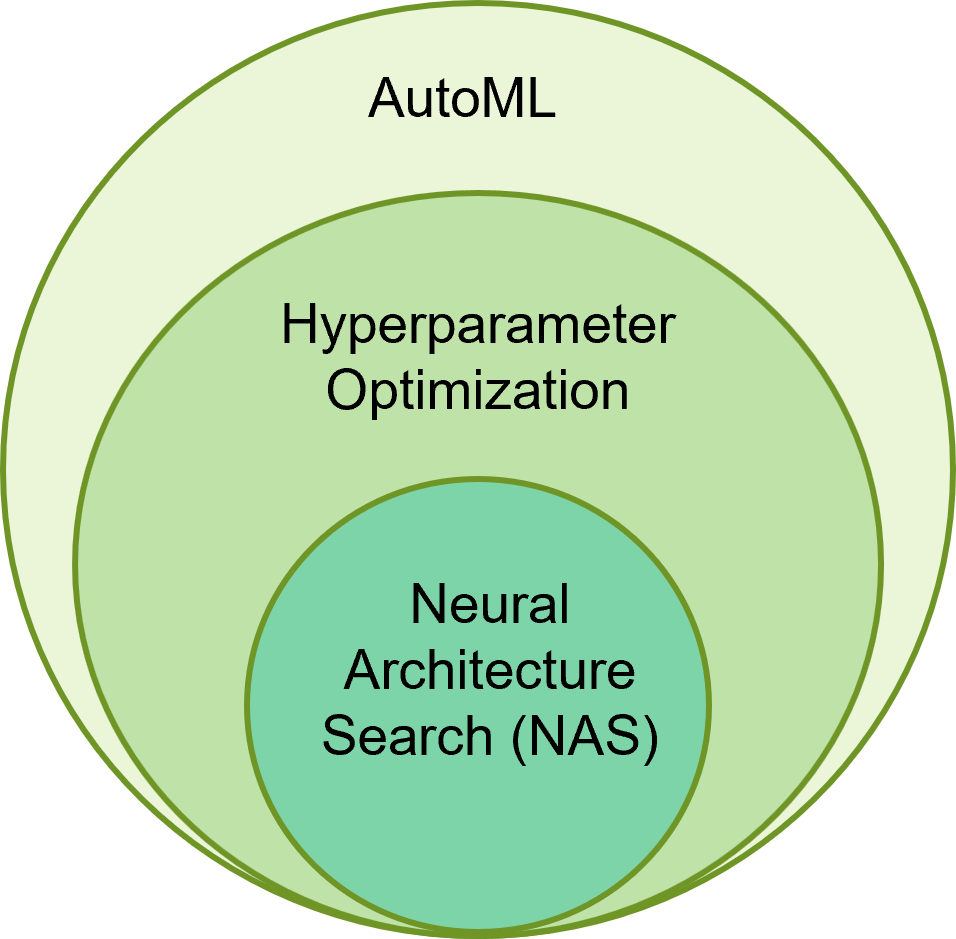}
\caption{ An overview of AutoML for discovering the most effective model through Neural Architecture Search (NAS)\cite{automl_NAS}}
\label{fig:automl_nas}
\end{figure}
\begin{figure}[t]
\centering
\includegraphics[scale=0.45]{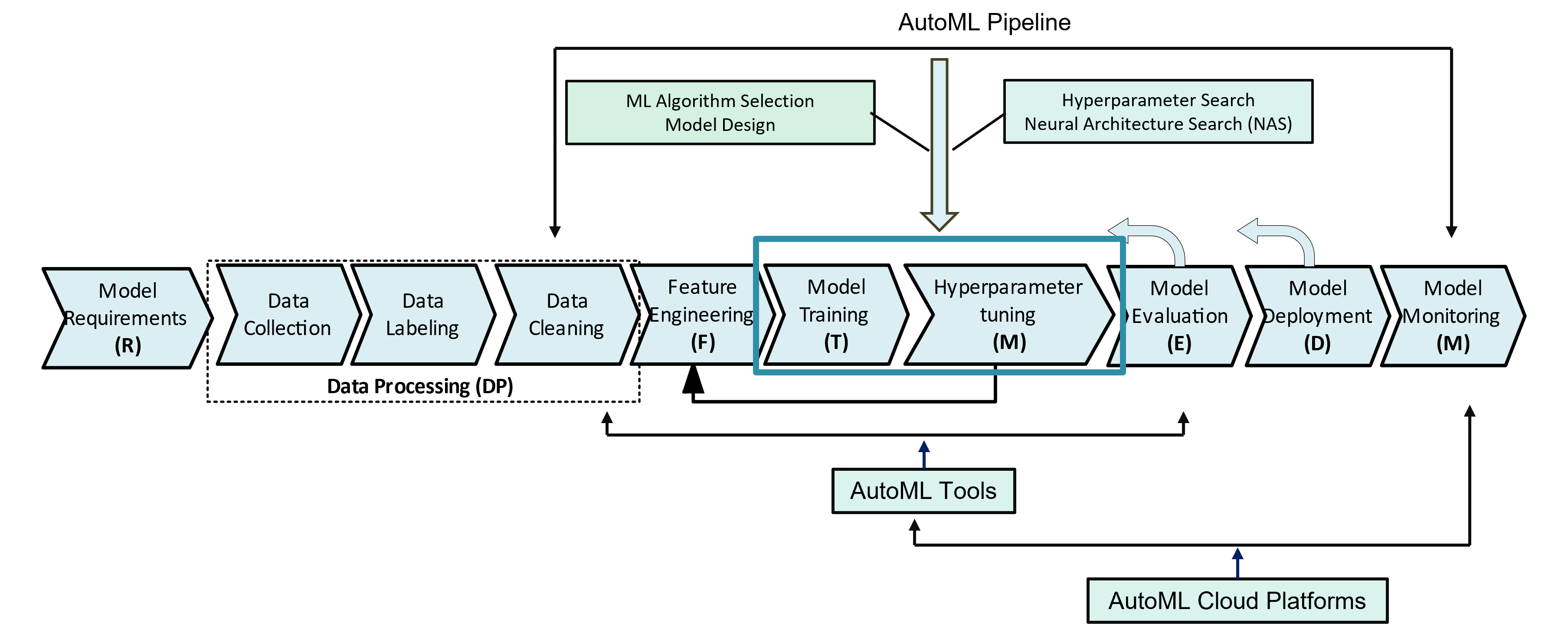}
\caption{ An overview of traditional ML pipeline vs AutoML pipeline.}
\label{fig:automl_pipeline}
\end{figure}

AutoML approach can be classified into two categories \begin{enumerate}
\item AutoML for traditional machine learning algorithms. It focuses on data pre-processing, feature engineering (i.e., finding the best set of variables and data encoding technique for the input dataset), ML algorithm section, and hyperparameter tuning.
\item AutoML for deep learning algorithms: this includes Neural architecture search (NAS), which generates and assesses a large number of neural architectures to find the most fitting one by leveraging reinforcement learning~\cite{sutton2018reinforcement} and genetic algorithm~\cite{mirjalili2019genetic}.
\end{enumerate}
Figure~\ref{fig:automl_nas} provides a high-level overview of AutoML for NAS and hyper-parameter optimization. The innermost circle represents the NAS exploration for the DL models, and the middle circle represents the hyper-parameter optimization search for both NAS and traditional ML applications.

\subsection{AutoML Services}
Depending on their mode of access/delivery, currently availalbe AutoML tools can be used over the cloud or via a stand-alone application in our desktop or internal server computers.

\bf{AutoML Cloud Service/Platforms.}
Large cloud providers and tech businesses started offering Machine Learning as a Service projects to make ML even more accessible to practitioners due to the rising popularity of AutoML tools. Some of these platforms specialize in various facets of AutoML, such as structured/unstructured data analysis, computer vision, natural language processing, and time series forecasting.
In 2016, Microsoft released AzureML~\cite{team2016azureml} runs on top of Azure cloud and assists ML researchers/engineers with data processing and model development.
H2O Automl~\cite{ledell2020h2o} was released in 2016, followed by H2O-DriverlessAI~\cite{h2odiverless} in 2017. It is a customizable data science platform with automatic feature engineering, model validation and selection, model deployment, and interpretability.
In 2017 Google released Google Cloud AutoML~\cite{googleautoml} that provides end-to-end support to train the custom model on the custom dataset with minimal effort.
Some other notable cloud platforms are Darwin (2018)~\cite{darwin} AutoML cloud platform for data science and business analytics, TransmogrifAI (2018)~\cite{TransmogrifAI} runs on top of Salesforce's Apache Spark ML for structured data.
This cloud-based AutoML platform enables end-to-end data analytics and AI solutions for nearly any sector.

\bf{AutoML Non-cloud Service (Tools/Library).}
The initial AutoML tools were developed in partnership with academic researchers and later by startups and large technology corporations~\cite{truong2019towards}. 
Researchers from the University of British Columbia and Freiburg Auto-Weka (2013)~\cite{kotthoff2019auto}, which is one of the first AutoML tools. 
Later, researchers from the University of Pennsylvania developed TPOT (2014)~\cite{olson2016evaluation}, and researchers from the University of Freiburg released Auto-Sklearn (2014)~\cite{feurer2020auto}. These three AutoML tools provide a higher level abstraction over the popular ML library ``SciKit-Learn'' (2007)~\cite{pedregosa2011scikit}. A similar research effort was followed to provide an automated ML pipeline over other popular ML libraries. University of Texas A\&M University developed Auto-Keras (2017)~\cite{jin2019auto} that provides runs on top of Keras~\cite{gulli2017deep} and TensorFlow~\cite{singh2020introduction}. Some of the other notable AutoML tools are MLJar (2018)~\cite{mljar}, DataRobot (2015)~\cite{datarobot},
tool named ``auto\_ml'' (2016)~\cite{auto_ml}. These AutoML tools provide a higher level of abstraction over traditional ML libraries such as TensorFlow, Keras, and Scikit-learn and essentially automate some ML pipeline steps (i.e., algorithm selection and hyper-parameter turning).

In Figure~\ref{fig:automl_pipeline}, we summarize the ML pipeline services offered by AutoML services. Traditional ML pipeline consists of various steps such as Model requirement, Data processing, feature engineering, model deigning, model evaluation, deployment, and monitoring~\cite{al2021quality, amershi2019software}. AutoML solutions aim to automate various stages of these pipelines, from data cleaning to Model deployment~\cite{truong2019towards} (Fig.~\ref{fig:automl_pipeline}). AutoML non-cloud solutions (i.e., tools/frameworks) mainly focus on automating different data filtering, model selection, and hyperparameter optimization. AutoML cloud platforms encapsulate the services of AutoML tools and provide model deployment and monitoring support. They usually provide the necessary tools for data exploration and visualization.

\section{Study Data Collection and Topic Modeling} \label{sec:methodology}

In this Section, we discuss our data collection process to find AutoML-related discussion, i.e., posts (Section~\ref{sub-sec:data_collection}). Then, we discuss in detail our data pre-processing and topic modeling steps on these posts (Section~\ref{sub-sec:topic_modeling}).

\subsection{Data Collection} \label{sub-sec:data_collection}
We collect  AutoML related SO posts in the following three steps: \begin{inparaenum}[(1)]
\item Download SO data dump,
\item Identify AutoML-related tag list, and
\item Extract AutoML-related posts using our AutoML tag list.
\end{inparaenum} We describe the steps in detail below.

\nd\textbf{Step 1: Download SO data dump.} For this study, we use the most popular Q\&A site, Stack Overflow (SO), where practitioners from diverse backgrounds discuss various software and programming-related issues~\cite{website:stackoverflow}. First, we download the latest SO data dump~\cite{SOdump} of June 2022, available during the start of this study. Following related studies~\cite{alamin2021empirical, iot21, abdellatif2020challenges}, we use the contents of the ``Post.xml'' file, which contains information about each post like the post's unique ID, title, body, associated tags, type (Question or Answer), creation date, favorite count, view-count, etc. Our data dump includes developers' discussions of 14 years from July 2008 to June 2022 and contains around 56,264,787 posts. Out of them, 22,634,238 (i.e., 40.2\%) are questions, 33,630,549 (i.e., 59.7\%) are answers, and 11,587,787 questions (i.e., 51.19\%) had accepted answers. Around 12 million users from all over the world participated in the discussions.


Each SO post contains 19 attributes, and some of the relevant attributes for this study are: \begin{inparaenum}[(1)]
        \item Post's unique Id, and creation time,
        \item Post's body with problem description and code snippets,
        \item Post's score, view, and favorite count,
        \item Tags associated with a post,
        \item Accepted answer Id.
    \end{inparaenum}

\nd\textbf{Step 2: Identify AutoML tags.}
We need to identify AutoML related SO tags to extract AutoML-related posts, i.e., practitioners' discussions. We followed a similar procedure used in prior work~\cite{alamin2021empirical, iot21,  abdellatif2020challenges, ahmed2018concurrency, wan2019discussed, linares2013exploratory} to find relevant SO tags. In Step 1, we identify the initial AutoML-related tags and call them $T_{init}$. In Step 2, we finalize our AutoML tag list following approaches of related work~\cite{bagherzadeh2019going, yang2016security}. Our final tag list $T_{final}$ contains 41 tags from the top 18 AutoML service providers. We discuss each step in detail below.

(1) Identifying Initial AutoML tags.
Following our related work~\cite{alamin2021empirical}, first, we compile a list of top  AutoML services. First, we make a query in google with the following two search terms ``top AutoML tools'' and ``top AutoML platforms''. We select the first five search results for each query that contain various websites ranked the best AutoML tools/platforms. The full list of these websites is available in our replication package. So, from these ten websites and the popular technological research website Gartner\footnote{https://www.gartner.com/reviews/market/data-science-machine-learning-platforms} we create a list of 38 top AutoML solutions. Then for each of these AutoML platforms, we search for SO tags. For example, We search for ``AutoKeras'' via the SO search engine. We find a list of SO posts discussing the AutoKeras tool. We compile a list of potential tags for this platform. For example, we notice most of these questions contain ``keras'' and ``auto-keras'' tags. Then, we manually examine the metadata of these tags \footnote{https://meta.stackexchange.com/tags}. For example, the metadata for the ``auto-keras'' tag says, ``Auto-Keras is an open source software library for automated machine learning (AutoML), written in python. A question tagged auto-keras should be related to the Auto-Keras Python package.'' The metadata for the ``keras'' tag says, ``Keras is a neural network library providing a high-level API in Python and R. Use this tag for questions relating to how to use this API. Please also include the tag for the language/backend ([python], [r], [tensorflow], [theano], [cntk]) that you are using. If you are using tensorflow's built-in keras, use the [tf.keras] tag.''. Therefore, we choose the ``auto-keras'' tag for the ``AutoKeras'' AutoML library. Not all AutoML platforms have associated SO tags; thus, they were excluded. For example, for AutoFolio~\cite{autofolio} AutoML library, there are no SO tags; thus, we exclude this from our list. This way, we find 18 SO tags for 18 AutoML services and call it $T_{init}$. The final AutoML solutions and our initial tag list are available in our replication package.

(2) Finalizing AutoML-related tags.
Intuitively, there might be variations to tags of 18 AutoML platforms other than those in $T_{init}$. We use a heuristic technique from related previous  works~\cite{alamin2021empirical, bagherzadeh2019going, yang2016security} to find the other relevant AutoML tags. First, we denote the entire SO data as $Q_{all}$. Second, we extract all questions $Q$ that contain any tag from $T_{init}$. Third, we create a candidate tag list $T_{candidate}$ using the relevant tags in the questions $Q$. Fourth, we analyze and select significantly relevant tags from  $T_{candidate}$ for our  AutoML discussions. Following related works~\cite{alamin_LCSD_EMSE, alamin2021empirical, iot21, bagherzadeh2019going, yang2016security}, we compute relevance and significance for each tag $t$ in $T_{candidate}$ with respect to $Q$ (i.e., the extracted questions that have at least one tag in $T_{init}$) and $Q_{all}$ (i.e., our data dump) as follows,
{ \[
( Significance) \ \ S_{tag} \ =\ \ \frac{\#\ of\ ques.\ with\ the\ tag\ t\ in\ Q}{\ \ \#\ of\ ques.\ with\ the\ tag\ t\ in\ Q_{all}}
\]

\[
( Relevance) \ \ R_{tag} \ =\ \ \frac{\#\ of\ questions\ with\ tag\ t\ in\ Q}{\ \ \#\ of\ questions\ in\ Q}
\]} A tag $t$ is significantly relevant to  AutoML if the $S_{tag}$ and  $R_{tag}$ are higher than a threshold value. 
Similar to related study\cite{alamin_LCSD_EMSE, iot21}, we experimented with a wide range of values of $S_{tag}$ = \{0.05, 0.10, 0.15, 0.20, 0.25, 0.30, 0.35\} and  $R_{tag}$ = \{0.001, 0.005, 0.010, 0.015, 0.020, 0.025, 0.03\}. 
From our analysis, we find that we increase $S_{tag}$ and  $R_{tag}$ the total number of recommend tags decreases. For example, we find that for $S_{tag}$ = .05 and  $R_{tag}$ = 0.001 the total number of recommended tags for AutoML is 30 which is highest. 
However, not all these recommended tags are AutoML-related. For example, ``amazon-connect'' tag's $S_{tag}$ = 0.24 and $R_{tag}$ = 0.006 and this tag is quite often associated with questions related to other AutoML tags such as ``aws-chatbot'', ``amazon-machine-learning'' but it mainly contains discussion related to AWS cloud based contact center solutions rather than AutoML related discussion. So, we remove this from our final tag list. Similarly, we find some other tags such as 
 ``splunk-formula'', ``amazon-ground-truth'' etc are frequently correlated with other AutoML platform tags, although they do not contain AutoML related discussions. After manually analysing these 31 tags we find that 23 new tags are relevant to AutoML-related discussions. 
So, after combining with out initial taglist, i.e., $T_{init}$, our final tag list $T_{final}$ contains 41 significantly relevant AutoML-related tags which are:
\begin{description}
\item Final Tag List $T_{final}$ = \{ `amazon-machine-learning', `automl', `aws-chatbot', `aws-lex', `azure-machine-learning-studio', `azure-machine-learning-workbench', `azureml', `azureml-python-sdk', `azuremlsdk', `driverless-ai', `ensemble-learning', `gbm', `google-cloud-automl-nl', `google-cloud-vertex-ai', `google-natural-language', `h2o.ai', `h2o4gpu', `mlops', `sparkling-water', `splunk-calculation', `splunk-dashboard', `splunk-query', `splunk-sdk', `amazon-sagemaker', `tpot', `auto-sklearn', `rapidminer', `pycaret', `amazon-lex', `auto-keras', `bigml', `dataiku', `datarobot', `google-cloud-automl', `h2o', `mljar', `splunk', `transmogrifai', `ludwig', `azure-machine-learning-service', `pycaret'\}
\end{description}

\nd\textbf{Step 3: Extracting AutoML related posts.}
Hence, our final dataset $B$ contained 14,341  posts containing 73.7\% Questions (i.e., 10,549 Q) and 26.3\% Accepted Answers (i.e., 3,792).

\subsection{Topic Modeling} \label{sub-sec:topic_modeling}
We produce  AutoML topics from the extracted posts in three steps: \begin{inparaenum}[(1)]
\item Preprocess the posts, 
\item Find the optimal number of topics, and
\item Generate topics.
\end{inparaenum} We discuss the steps in detail below.

\nd\textbf{Step 1. Preprocess the posts.} For each post text, we remove noise using the technique in related works~\cite{alamin_LCSD_EMSE, alamin2021empirical, iot21, abdellatif2020challenges,bagherzadeh2019going, barua2014developers}. First, we remove the code snippets from the post body, which is inside \textless code\textgreater \textless /code\textgreater\ tag, HTML tags such as (\textless p\textgreater \textless /p\textgreater, \textless a\textgreater \textless /a\textgreater, \textless li\textgreater \textless /li\textgreater\, etc.), and URLs. Then we remove the stop words such as ``am'', ``is'', ``are'', ``the'', punctuation marks, numbers, and non-alphabetical characters using the stop word list from MALLET~\cite{mccallum2002mallet}, NLTK~\cite{loper2002nltk}.
After this, we use porter stemmer~\cite{ramasubramanian2013effective} to get the stemmed representations of the words, e.g., ``waiting'', ``waits'' - all of which are stemmed to the base form of the word ``wait''.

\nd\textbf{Step 2. Finding the optimal number of topics.}  After the prepossessing, we use Latent Dirichlet Allocation~\cite{blei2003latent} and the MALLET tool~\cite{mccallum2002mallet} to find out the AutoML-related topics in our SO discussions. We follow similar studies using topic modeling~\cite{alamin_LCSD_EMSE, alamin2021empirical, iot21, arun2010finding, asuncion2010software, yang2016security, bagherzadeh2019going,abdellatif2020challenges} in SO dataset. Our goal is to find the optimal number of topics $K$ for our AutoML dataset $B$ so that the \textit{coherence} score is high, i.e., encapsulation of underlying topics is more coherent. We use Gensim package~\cite{rehurek2010software} to determine the coherence score following previous research~\cite{alamin2021empirical, uddin2017automatic, roder2015exploring}. We experiment with different values of $K$ that range from \{5, 10, 15, 20, 25, 30, 35, 40, 45, 50, 55, 60, 65, 70\} and for each value, we run  MALLET LDA on our dataset for 1000 iterations~\cite{alamin_LCSD_EMSE, iot21, bagherzadeh2019going}. Then we observe how the coherence score changes with respect to $K$. As LDA topic modeling has some inherited randomness, we ran our experiment 3 times and found that we got the highest coherence score for $k$ = 15. Choosing the right value of $K$ is crucial because multiple real-world topics merge for smaller values of $K$, and for a large value of $K$, a topic breaks down. MALLET uses two hyper-parameters, $\alpha$, and $\beta$, to distribute words and posts across the generated topics. Following the previous works~\cite{alamin2021empirical, iot21, bagherzadeh2019going, ahmed2018concurrency, bajaj2014mining, rosen2016mobile}, in this study we use the standard values $50/K$ and 0.01 for hyper-parameters $\alpha$ and $\beta$ in our experiment.  

\nd\textbf{Step 3. Generating topics.} Topic modeling is a systematic approach to extracting a set of topics by analyzing a collection of documents without any predefined taxonomy. Each document, i.e., the post, has a probability distribution of topics, and every topic has a probability distribution of a set of related words~\cite{barua2014developers}. We generate 15 topics using the above LDA configuration on our AutoML dataset $B$.
Each topic model provides a list of the top $N$ words and a list of $M$ posts associated with the topic. A topic in our context comprises the 30 most commonly co-related terms, which indicate an AutoML development-related concept. Each post has a correlation score between 0 to 1, and following the previous work~\cite{alamin_LCSD_EMSE, alamin2021empirical, iot21,  wan2019discussed}, we assign a document, i.e., a post with a topic that corresponds most.

\section{Empirical Study} \label{sec:results}
We answer the following four research questions by analyzing the topics we found in our 14.3K AutoML related posts in Stack Overflow (SO).
\begin{enumerate}[label=\bf{RQ\arabic{*}.}]
    \item What topics do practitioners discuss regarding AutoML services on SO?
    \item How are the AutoML topics distributed across machine learning life cycle phases?
    \item What AutoML topics are most popular and difficult on SO?
    \item How does the distribution of the topics differ between cloud and non-cloud AutoML services?
\end{enumerate}


\subsection{RQ1. What topics do practitioners discuss regarding AutoML services on SO?}\label{rq:topic}
\subsubsection{Motivation}
AutoML is increasingly gaining popularity for domain experts to develop machine learning applications without requiring a comprehensive understanding of the machine learning process. The challenges posed by AutoML tools/platforms needs to be investigated as a distinct strategy for ML application development. SO is a popular discussion platform for systematically analyzing and studying practitioners' challenges. A detailed analysis of the practitioners' challenges in SO will provide invaluable insight to AutoML researchers and providers into the existing obstacles to AutoML adoption. This analysis will also provide direction for future initiatives to mitigate these obstacles and make machine learning more accessible for various applications and research problems.

\subsubsection{Approach} \label{sub-sec:lda-labeling}
We apply LDA topic modeling to our AutoML-related discussion in SO. We get 13 AutoML-related topics, as discussed in \sec\ref{sec:methodology}. Following previous works ~\cite{bagherzadeh2019going, ahmed2018concurrency, yang2016security, rosen2016mobile, abdellatif2020challenges}, we use open card sorting~\cite{fincher2005making} approach to label these AutoML topics. 
In this approach, there is no predefined list of labels. Following related works~\cite{alamin2021empirical,iot21,bagherzadeh2019going,abdellatif2020challenges}, we label each topic by analyzing the top 30 words for the topic and a random sample of at least 20 questions from each topic. We assign a label to each topic and discuss it until there is an agreement.

After this initial analysis, i.e., labeling, we merge two topics if they contain similar discussions. For example, we merge topics 1 and 3 into Library/Platform Management because both contain developers' discussions about setting up the development environment and installing packages in a local or cloud environment. 
Similarly, we merge topics 5 and 13 into Data Management because they contain discussion related to data mining, data storing, and data filtering. In the end, we get 13 different AutoML-related topics.

After this, we reexamined the topic to find any clusters/groups. For example, Data Exploration, Data Management, and Data Transformation topics are related, and thus, they are grouped under the Data topic category. Similarly, we put Model Training \& Monitoring, Model Debugging, and Model Design topics under the Model topic category. We use an annotation guideline for labeling the topics and creating the taxonomy of topics to ensure consistency and reproducibility. We share the annotation guide with our replication package.

\subsubsection{Result} \label{rq_result:topics}
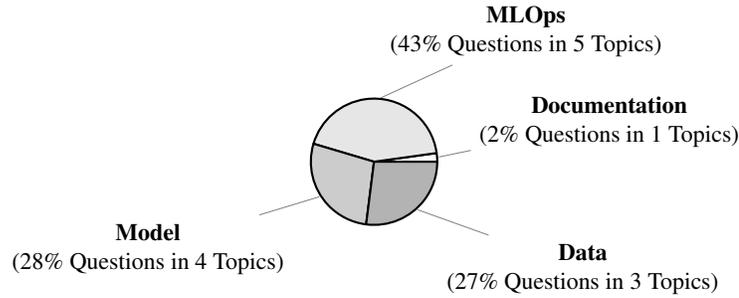
\begin{figure}[t]
	\centering\begin{tikzpicture}[scale=0.28]-
    \pie[
        /tikz/every pin/.style={align=center},
        text=pin, number in legend,
        explode=0.0,
        color={black!0, black!10, black!20, black!30, black!40},
        ]
        {
            2.2/\bf{Documentation} \\ (2\% Questions in 1 Topics),
            43.2/\bf{MLOps} \\ (43\% Questions in 5 Topics),
            27.6/\bf{Model} \\ (28\% Questions in 4 Topics),
            27/\bf{Data} \\ (27\% Questions in 3 Topics)
        }
    \end{tikzpicture}
	\caption{Distribution of Questions and Topics per Topic Category.}
	\vspace{-5mm}
	\label{fig:distribution_of_questions_topic_categories_pie_chart}
\end{figure}

After manually labeling and merging topics, we find 13 AutoML-related topics. Then after grouping these 13 topics into categories, we find four topic categories: \begin{inparaenum}[(1)]
\item MLOps, 
\item Model, 
\item Data, and
\item Documentation
\end{inparaenum}. Figure \ref{fig:distribution_of_questions_topic_categories_pie_chart} shows the distribution of these four topic categories and their percentage of questions. Among these categories, MLOps has the highest coverage of questions and topics (43\% Questions in 5 Topics), followed by the Model category (28\% Questions in 4 Topics), Data (27\% Questions in 3 Topics), Documentation (2\% Questions in 1 Topic).

\begin{figure}[t]
\centering
\includegraphics[scale=0.55]{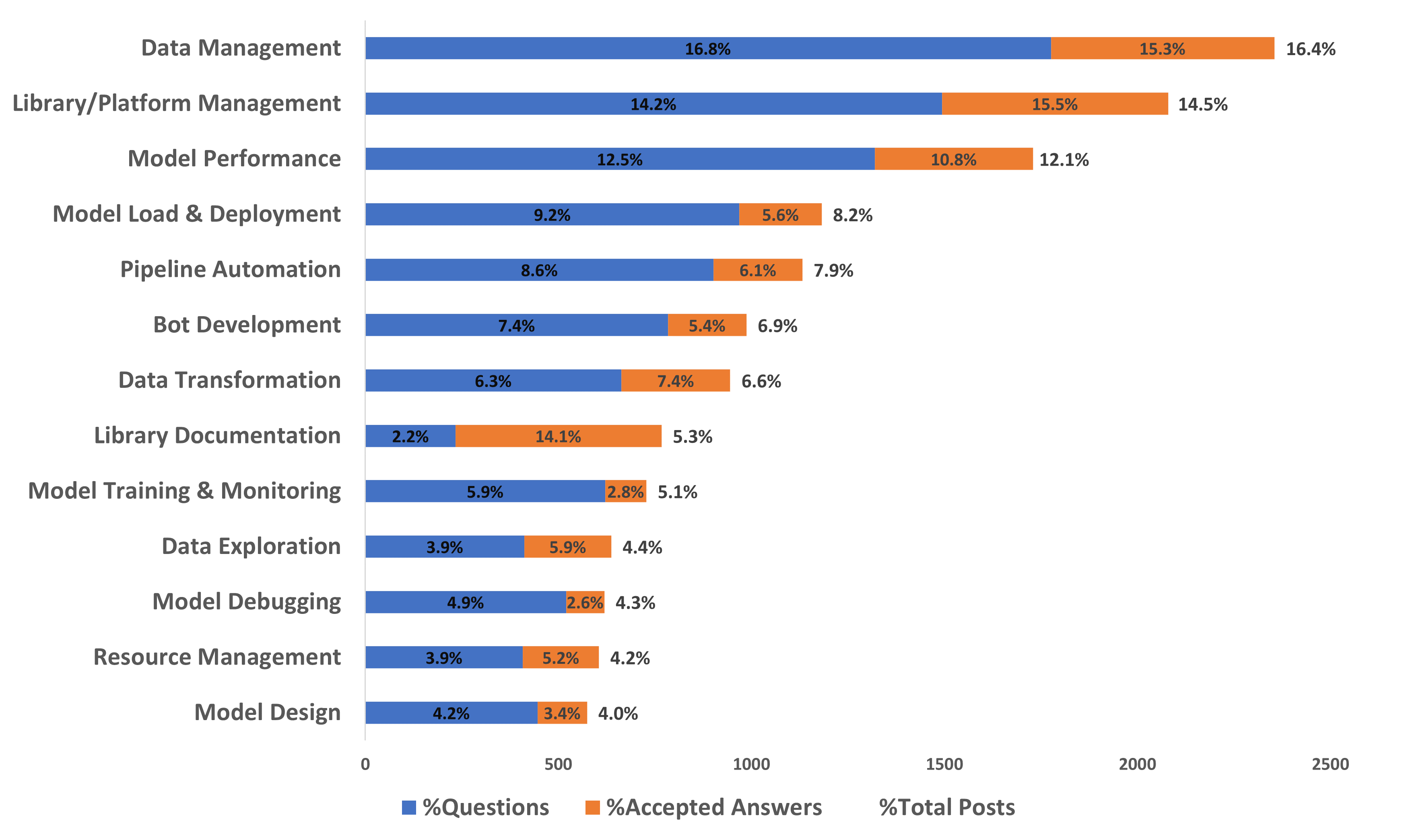}
\caption{Distribution of posts (\#Questions + \#Accepted Answers) of AutoML topics.}
\label{fig:distribution_of_topic_posts_bar_chart}
\end{figure}

Figure~\ref{fig:distribution_of_topic_posts_bar_chart} presents the 13 AutoML topics sorted by the number of posts. A post is either an AutoML-related question or an accepted answer. As discussed in Section \ref{sub-sec:data_collection}, our final dataset contains 14,341 posts containing 10,549 questions and 3,792 accepted answers. The topic with the highest number of posts is at the top of the list. On average, each AutoML topic has 1103 posts (\#question + \#accepted answer). The topic Data Management has the highest number of posts (16.4\%), containing 16.8\% of total questions and 15.3\% total accepted answers. On average, each topic has around 811 AutoML-related SO questions.

\begin{figure}[t]
\centering
\includegraphics[scale=0.65]{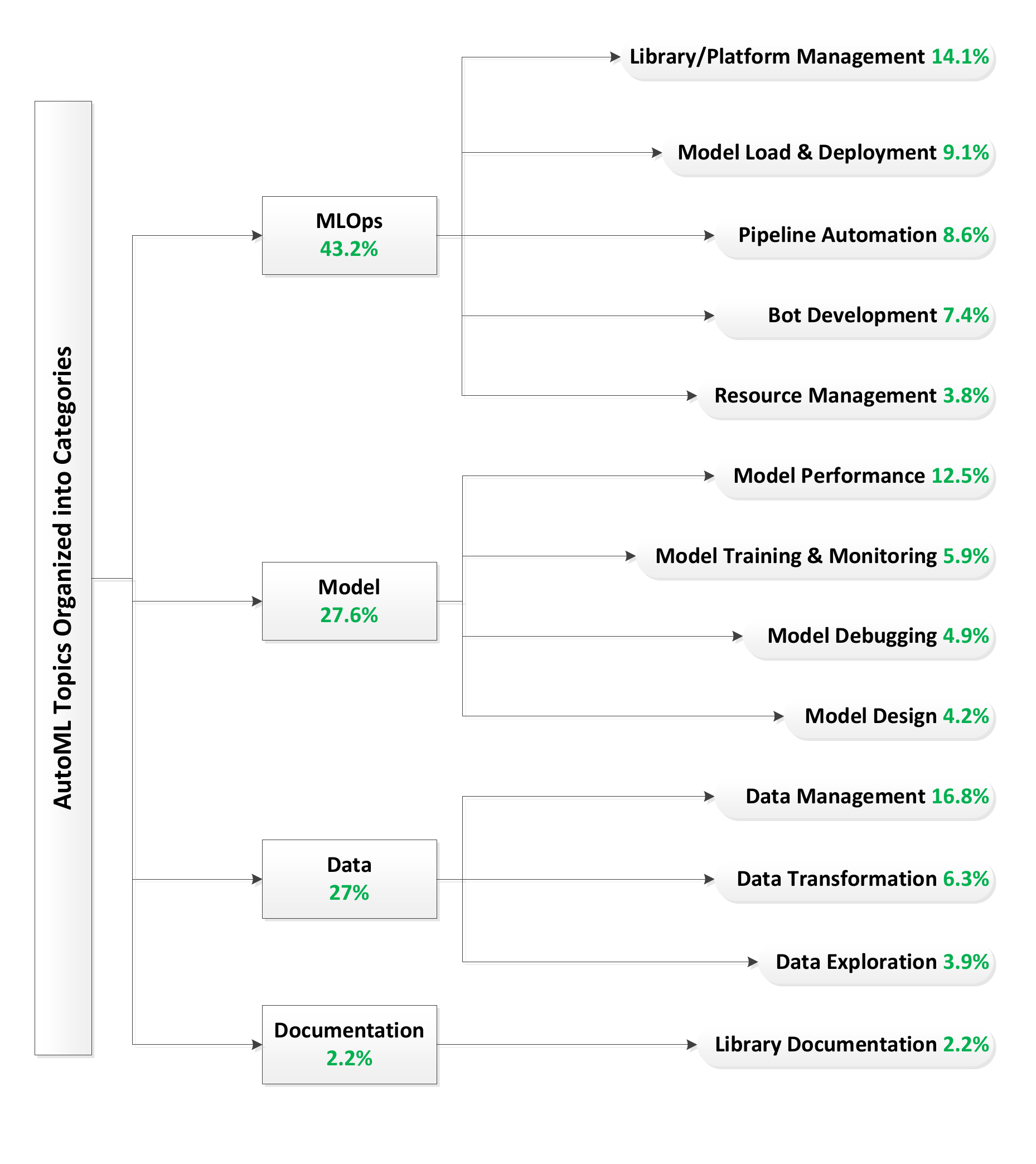}
\caption{A taxonomy of 13 AutoML topics organized into four categories.}
\label{fig:taxonomy_TM}
\end{figure}

Figure \ref{fig:taxonomy_TM} describes 13 AutoML-related topics into four categories. The topics are presented in descending order of the number of questions. For example, the MLOps category has the highest number of questions, followed by the Model category. Similarly, topics under each category are also organized in descending order of the number of questions. For example, Under the Data topic category, three topics: Data Management (16.8\% questions), Data transformation (6.3\% questions), and Data exploration (3.9\% questions), are organized by the number of questions.
We discuss the four AutoML topic categories and the 13 topics below.

\nd\bf{\ul{MLOps}} is biggest topic category based on number of questions (43.2\% questions). MLOps~\cite{mlops} is a term for Machine Learning Operations, and it focuses on practices to streamline the process of ML engineering from model development to production to monitor their performance over time. This collaborative approach involves stakeholders from different teams, such as ML engineers and DevOps engineers. This category discusses automating the machine learning pipeline using the AutoML services, setting up local and cloud environments, maximizing resource (i.e., CPU, GPU) management, and deploying the trained model. It has five topics:
\begin{inparaenum}[(1)]
    \item Topic \textit{Library/Platform Management (14.1\%)} contains discussion about setting up a proper development environment, installing (\dq{51471526}), upgrading, and downgrading packages (\dq{70047920}), resolving incompatible version issues (\dq{53328131}, \dq{71569924}), portability issues in different operating systems (\dq{67940893}) both for local and cloud platforms. For example, in \dq{56215238}, a user is inquiring about integrating a third-party TensorFlow library with H2O's flow UI (e.g., ``Install new package in H2O Flow UI'').
    \item Topic \textit{Model Load \& Deployment (9.1\%)} topic contains discussion related to training and saving the trained model (\dq{60733257}), deploying the trained model via an API endpoint for public use (\dq{71513161}, \dq{69338516}), how to export a trained host locally (\dq{56842056}, \dq{64467754}) or different cloud environment (\dq{54145693}). For example, in \dq{71513161}, a practitioner is asking for help to deploy the newly trained object detection model via AWS Sagemaker public endpoint (e.g., ``How can I create an endpoint for yolov5 inference in AWS sagemaker'')
    \item Topic \textit{Pipeline Automation (8.6\%)} contains discussion about different platform/tool specific features to automating different ML pipelines (\dq{68115476}), automatically scheduling experiments (\dq{30092360}), custom script to connect and extract data (\dq{59445324}), automating data pipeline(\dq{42651900}), automating CI/CD (\dq{67945601}), running profiler (\dq{61736810}). For example, this topic contains queries like ``How to organize one step after another in [AutoML Platform] Pipelines?'' (in \dq{68115476}).
    \item Topic \textit{Bot Development (7.4\%)} contains discussions about how to design, develop and improve the performance of different bots (\dq{46886079}), different challenges of developing chatbot (\dq{55708995}, \dq{69490831}, \dq{52439304}), i.e., managing different intents (\dq{64008462}), integrating existing bots such as Alexa (\dq{50993053}) with new applications, integration of different bots (\dq{71675543}), bot deployment (\dq{71220465}). So, in this topic, practitioners ask queries such as ``Approaches to improve Microsoft ChatBot with each user conversation by learning from it?'' in \dq{46886079} or ``How to integrate Amazon lex with MS Bot framework?'' in \dq{71675543}. \textit{We put this topic under MLOps rather than the Model category because after analyzing, we find most of the questions of this topic mostly focus on streamlining ML operations by integrating various cloud services, as opposed to the training or debugging the model.}
    \item Topic \textit{Resource Management (3.8\%)} contains discussions related to resource management for AutoML tools or better configuring the AutoML cloud platform's training setup. It mainly concerns queries regarding how to maximize GPU utilization (e.g., ``Can Driverless AI instance utilize K80 GPU for faster performance in GCP?'' in\dq{64651996}, \dq{60243228}), CPU utilization (\dq{71121820}, \dq{36680046}), using optimal containers/nodes (\dq{65276017}), Auto scaling (e.g., ``how does scaling policy work with sagemaker endpoints?'' in \dq{71344215}, \dq{69904211}).
\end{inparaenum}

\nd\bf{\ul{Model}} is the second largest topic category based on the number of questions (27.6\% questions). It discusses designing ML models, training, debugging and improving performance. It has four topics:
\begin{inparaenum}[(1)]
    \item Topic \textit{Model Performance (12.5\%)} contains discussion related to improving the AutoML model's performance, i.e., queries related to why one model performs better than another (\dq{62167244}), how to improve performance (e.g., ``how to improve gradient boosting model fit'' in \dq{61002095}), compare, monitor and visualize performance metrics (e.g., ``How do you monitor more 'standard' metrics with SageMaker?'' in \dq{71898075}) using the AutoML dashboard.
    \item Topic \textit{Model Training \& Monitoring (5.9\%)} contains discussions about different issues with training the model (e.g., ``Why is my Google automl vision trained on cloud much better than the one trained on edge'' in \dq{59896056}), deleting trained model (\dq{57750611}), programmatically training the model (\dq{60329106}, \dq{51873458}), resources it takes for training (60535298), explainability of the model (e.g., ``Explainable AI with Cloud Vision API'' in \dq{72127541}, \dq{68251877}), integrate the model with mobile/web application (\dq{54781708}, \dq{72081132}), monitor the training progress (e.g., ``Vertex AI - how to monitor training progress?'' in \dq{72051655}).
    \item Topic \textit{Model Debugging (4.9\%)} contains discussion related to the implementation of different ML algorithms and debugging queries such as importing library (\dq{37549591}), error messages (e.g., ``H2OAutoML throws libgomp exception in train step'' in \dq{57758996}, \dq{66011434}, \dq{61803264}), debugging invalid arguments (e.g., ``Invalid argument to unary operator'' in \dq{41058261}), debugging codes (\dq{65812334}).
    \item Topic \textit{Model Design (4.2\%)} contains discussions on the theoretical discussion of ML experiment design using AutoML services. It contains questions requesting suggestions for the best approach for using AutoML platforms/tools for regression or classification tasks. For example, it contains questions related to machine learning experiment design (\dq{12651719}), such as which ML algorithm and metrics to use for various problems (e.g., ``Which machine learning algorithm to evaluate the best combinations of groups?'' in \dq{43332966}, \dq{37966070}), feature engineering (e.g., ``Questions about feature selection and data engineering when using H2O autoencoder for anomaly detection'' in \dq{64490264}),  time series forecasting (\dq{54862099}), unsupervised learning (e.g., ``Unsupervised Sentiment Analysis pycaret'' in \dq{63143622}).
\end{inparaenum}

\nd\bf{\ul{Data}} is the third largest topic category based on the number of questions (27\% questions). It contains practitioners' discussion on data collection and management, data filtering and exploration, and transforming and querying the data. It has three topics:
\begin{inparaenum}[(1)]
    \item Topic \textit{Data Management (16.8\%)} is the biggest topic by the number of questions, and it contains discussion related to the dashboard to configure and manage data mining and security setup (e.g., ``Splunk dashboard and reports source backup for versioning'' in \dq{68821885}, \dq{61658701}), automatically sending data for mining (e.g., ``How do I send JSON files to Splunk Enterprise from JAVA?'' in \dq{58555219}, \dq{23017469}) and reporting the mining result to another application (\dq{57149307}) or via email (\dq{58204476}), 
    \item Topic \textit{Data Transformation (6.3\%)} contains discussions related to data encoding, i.e., convert to categorical features (\dq{50987850}, \dq{47513901}), processing data frames (e.g., ``how to join two frames in h2o flow?'' in \dq{37344958}, \dq{67452392}, \dq{56246048}), data filtering (e.g., ``Filtering h2o dataset by date, but being column imported as time in R'' in \dq{51216030}), finding patterns using regex (\dq{10890718}), read and process large data files (e.g., ``How to read large text file stored in S3 from sagemaker jupyter notebook?'' in \dq{65550808}),  processing millions of files (\dq{62330719})
    \item Topic \textit{Data Exploration (3.9\%)} contains discussions related to creating new attributes in the dataset (\dq{26195630}), making/storing/exporting data association rule (e.g., ``Make association rules of a certain template'' in \dq{30684638}, \dq{52568212}), data pre-processing (\dq{25056600}) and mining (e.g., ``Text Mining a single text document'' in \dq{22885133}, \dq{24241992}), issues with making queries on the dataset (e.g., ``MongoDB aggregation query in RapidMiner'' in \dq{41068724}, \dq{15229636}),
\end{inparaenum}

\nd\bf{\ul{Documentation}} is smallest topic category based on number of questions (2.2\% questions). It contains a discussion related to various limitations on AutoML API documentation and instructions on using other libraries. It has one topic:
\begin{inparaenum}[(1)]
    \item Topic \textit{Library Documentation (2.2\%)} contains discussion about AutoML tool/platforms API documentation. For example, it contains questions inquiring about resources for details about API parameters (e.g., requesting details on documentation ``Python: h2o detailed documentation'' in \dq{60667920}, \dq{72385022}, \dq{52305760}), API constraint/unclear API description (e.g., ``Could feature interaction constrains be enforced in h2o.xgboost in R'' in \dq{64247245}), Incorrect API documentation, requesting examples for API usage (e.g., ``What are some specific examples of Ensemble Learning?'' in \dq{38135557}, \dq{31172775}).
\end{inparaenum}

\begin{tcolorbox}[flushleft upper,boxrule=1pt,arc=0pt,left=0pt,right=0pt,top=0pt,bottom=0pt,colback=white,after=\ignorespacesafterend\par\noindent]
\noindent\textbf{RQ1. What topics do practitioners discuss regarding AutoML services on SO?}
We found 13 AutoML-related topics that we organized into four categories. The MLOps category (5 topics) has the highest number of SO questions (43.2\%), followed by Model (4 topics, 27.6\% questions), Data (3 topics, 27\% questions), documentation (1 topic, 2.2\% questions). We find that the Data Management topic under the Data category has the highest number of questions (16.8\%), followed by Library/Platform Management (14.2\%) under MLOps, Model Performance (12.5\%) under Model. Our study reveals that AutoML practitioners mostly struggle with integrating different AutoML services to improve model performance and setting up development and deployment environment.
\end{tcolorbox}
\subsection{RQ2. How AutoML topics are distributed across different machine learning life cycle phases?}\label{rq:ml_sdlc}

\subsubsection{Motivation}
From our previous analysis, we see that the AutoML topics contain different types of questions. We also observe that these topics are asked about different Machine Learning Life Cycle (MLLC) phases. Machine learning is a distinct domain with its life cycle phases. In order to complete an ML application, a practitioner has to go through an MLLC phase such as problem defining, data collection and processing, model development, deployment, and monitoring. For example, API documentation-related questions are mostly asked during the Model Training phase. Therefore an in-depth analysis of questions asked during the six MLLC phases may offer us a better understanding of the discussed AutoML challenges. In this research question, we aim to investigate which ML phases are most challenging and how these topics are distributed across these MLLC phases. This analysis may provide AutoML practitioners with a better understanding of ML application development using AutoML tools/platforms. This analysis will provide valuable insights for AutoML researchers and tools/platform providers.

\subsubsection{Approach}\label{sub-sec:mllc}

In order to better understand how the questions are discussed in SO on MLLC, we take a statistically significant sample from our extracted dataset and then manually analyze the questions and label them into one of MLLC phases. So, our approach is divided into two main steps: Step 1. We generate a statistically significant sample size from our dataset, Step 2. we manually analyze and label those questions.

\nd \textbf{Step 1. Generate Sample.} As we discussed in Section \ref{sub-sec:data_collection} our dataset has 10,549 SO questions. A statistically significant sample with a 95\% confidence level and five confidence intervals would be at least 371 random questions, and a 10 confidence interval would have a sample size of 95 questions. 
A random sample is representative of the entire dataset, but it could miss questions belonging to smaller topic categories. For example, as discussed in RQ1, we have 13 AutoML topics organized into four categories. As random sampling from the dataset is not uniform across the topic categories, it might miss important questions from smaller topic categories such as Documentation. Therefore, following previous empirical studies \cite{iot21, abdellatif2020challenges}, we extract a statistically significant random sample question from each of the four topic categories. More specifically, we extract SO questions in our sample from each of the four topic categories with a 95\% confidence level and 10 confidence intervals. The sample dataset is drawn as follows:  94 questions from the MLOps topic category (total question 4,557), 93 questions from the Model topic category (total question 2,907), 93 questions from the Data topic category (total question 2,851), 68 questions from Documentation topic category (total question 234). In summary, we sample a total of 348 questions.

\nd \textbf{Step 2. Labeling MLLC phase.} We analyze and label questions from our samples into the following six phases. Following similar work by Alshangiti et al.~\cite{alshangiti2019developing} and others~\cite{al2021quality, islam2019developers}, we label each question into one of the following six phases. \begin{inparaenum}[(1)]
\item Requirement Analysis, 
\item Data Preparation, 
\item Model Designing, 
\item Training, 
\item Model Evaluation, 
\item Model Deployment \& Monitoring.
\end{inparaenum}


\subsubsection{Result} \label{rq_result:mllc}

\begin{figure}[t]
	\centering
	\resizebox{4.8in}{!}{
	\begin{tikzpicture}
    \pie[
        explode=0.0, text=pin, number in legend, sum = auto, 
        color={black!0,black!10, black!20,black!30,black!40,black!60},
        ]
        {            
            28.7/\LARGE{Training} 28.7\%,
            27.7/\LARGE{Data Preparation} 24.7\%,
            18.7/\LARGE{Model Designing} 18.7\%,
            12.9/\LARGE{Model Deployment \& Monitoring} 12.9\%,
            9.2/\LARGE{Model Evaluation} 9.2\%,
            5.7/\LARGE{Requirement Analysis} 5.7\%
        }
    \end{tikzpicture}
    }
	\caption{Distribution of questions (Q) per MLLC phase}
	\label{fig:distribution_of_SDLC_pie_chart}
\end{figure}
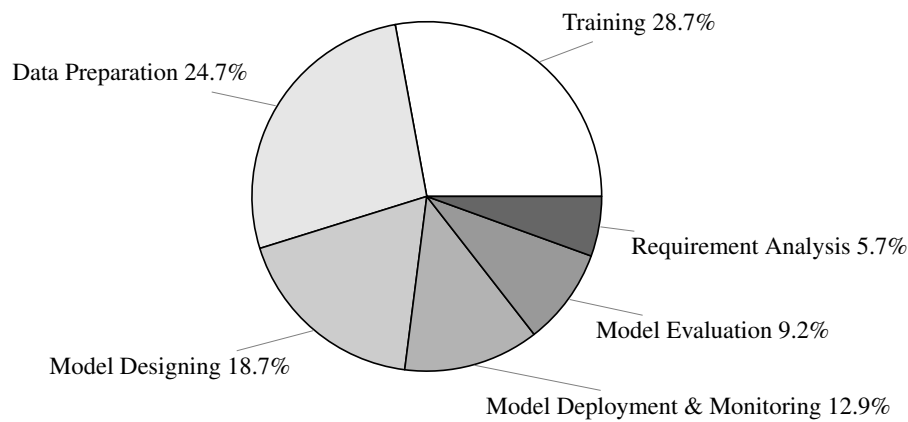

\begin{table}[t]
  \centering
   \caption{Distribution (frequency) of AutoML topics per MLLC phase. Each colored bar denotes a phase 
   (Black = Requirement Analysis, Green = Data Preparation, Magenta = Model Designing, Red = Training, Blue = Model Evaluation, Orange = Model Deployment \& Monitoring)}
    \begin{tabular}{lr}
    \toprule{}
    \textbf{Topics} & \textbf{MLLC Phases Found in \#Questions}\\
    \midrule
        \textbf{MLOps (94)} & \sixbars{4}{9}{10}{37}{8}{26}  \\
        \textbf{Model (93)} & \sixbars{7}{9}{27}{24}{18}{8}  \\
        \textbf{Data (93)} & \sixbars{2}{59}{11}{11}{2}{8} \\
        \textbf{Documentation (68)} & \sixbars{7}{10}{17}{27}{4}{3}  \\
    \bottomrule
    \end{tabular}%
  \label{tab:topicSDLC}%
\end{table}%

Figure~\ref{fig:distribution_of_SDLC_pie_chart} presents the distribution of our AutoML questions into six MLLC phases. We find that the Model training, i.e., the implementation and debugging phase, has 28.7\% of our 348 annotated questions, followed by Data Preparation (24.7\%), Model Designing (18.7\%), Model Deployment \& Monitoring (12.9\%), Model Evaluation (9.2\%), and Requirement Analysis (5.7\%). SO is a technical Q\&A platform, and developers/practitioners use it for solving technical issues; thus, we find many debugging and troubleshooting related queries in Model implementation and data pre-processing related. However, we also find that AutoML practitioners ask many queries related to designing (e.g., \dq{56086214}) and deploying AutoML applications (e.g., \dq{69113428}). This analysis highlights that AutoML practitioners struggle in using AutoML services.
We provide an overview of the types of questions asked during these six MLLC phases.

\bf{Requirement Analysis (20, 5.7\%).}
This is the first stage of any ML project. In this phase, the scope and definition of the problem are defined based on the project's goal. There are questions on the AutoML tools/platforms' features (e.g., ``Does AzureML RL support PyTorch?'' in \dq{64025308}), architectural design of the platform (e.g., ``What is Splunk architecture?'' in \dq{28186144}, comparison of different AutoML services (e.g., ``What are the differences between AWS sagemaker and sagemaker pyspark?'' in \dq{66817781}), condition for their usage (e.g., ``Is it possible to share compute instance with other user?'' in \dq{60539094}).

\bf{Data Preparation (86, 24.7\%).} 
Based on the problem definition, necessary data is collected from many different sources. The performance of the ML model largely depends on the quality of the data. So proper data filtering (e.g., \dq{55753324}), pre-processing, formatting, visualization (e.g., in \dq{62188204}), and transformation is very important. So, in this phase, we find questions related to large data processing (e.g., ``Reading a large csv from a S3 bucket using python pandas in AWS Sagemaker'' in \dq{48111034}).

\bf{Model Designing (65, 18.7\%).}
At this step, a supervised or unsupervised learning approach is applied to the collected dataset for various tasks such as regression, classification, clustering, and forecasting. At this step, the optimal set of features from the dataset is chosen for the ML algorithms. This crucial step determines the performance of the prediction of the model. So, in this phase, we see discussion related to different theoretical discussions of different approaches for the ML model (e.g., in \dq{26410001}). For example, in \dq{15342081}, a practitioner is querying about how to use `:)' and `:(' emoticons to classify the sentiment of sentences (e.g., ``sentiment analysis for arabic language''). We can also see queries regarding the implementation of custom algorithms in AutoML platforms (e.g., in \dq{56086214})

\bf{Model Training (100, 28.7\%).}
At this stage, the implementation of the ML model and training takes place. At this step, the model learns from data and gets better at prediction. So, in this phase, we find many queries related to different issues of model training (e.g., ``Cannot run huge Python program'' in \dq{35237226}), debugging different issues, using AutoML cloud platforms UI for training (e.g., ``Amazon Lex slot won't save value in it'' in \dq{64244554}), etc.

\bf{Model Evaluation (32, 9.2\%).} 
At this stage, the performance of the trained model is evaluated on unseen data and reported using different metrics. If the output is not desirable, then the previous steps are iterated. So, in this phase, we find queries related to the prediction of the trained model. For example, in \dq{57858737}, a practitioner is querying about different predictions with the same model in the cloud vs. the local environment. Practitioners also ask questions related to error while making a prediction (e.g., ``Error in a multiclass classification in during of a prediction (Pycaret Predict function)'' in \dq{70956772}).

\bf{Model Deployment \& Monitoring (45, 12.9\%).}
This is the last stage of an ML application. The trained model is deployed into the production environment for general usage at this phase. Before deploying the model, the robustness and scalability need to be considered. The deployed model must be consistently monitored to ensure its prediction on the unseen dataset. At this step, we find queries related to deploying models in the AutoML cloud environment (e.g., ``Model deployment in Azure ML'' in \dq{69113428}), monitoring the model's prediction (e.g., \dq{71471108}), exporting the trained model from the cloud environment (e.g., \dq{56952741}).

\nd\bf{\ul{Topic Categories in different MLLC phases.}}
We find that for all four topic categories, AutoML practitioners need some community support from requirement analysis to implementation to deployment (e.g., ``Deploying multiple huggingface model through docker on EC2'' in \dq{71187430}). 
We analyze and present how AutoML topics are distributed across six MLLC phases. Table~\ref{tab:topicSDLC} presents the distribution of six MLLC phases for each topic category. Our analysis shows that the MLOps topic Category, which encapsulates overall machine learning operations from model development to deployment, contains questions that are asked during the Model development and training (39\%) and Deployment (28\%) phases regarding various MLOps. Similarly, most questions from Model topic categories are dominant during the model design (29\%) and model training phase (26\%) regarding how to best design and train the ML model. Around 63\% questions from the Data topic category are asked during the data preparation and pre-processing phase regarding data filtering and data querying. We also find that the Document topic category is mostly asked during model training (25\%) and model designing phase (40\%), highlighting the different limitations of API documentation.


\begin{tcolorbox}[flushleft upper,boxrule=1pt,arc=0pt,left=0pt,right=0pt,top=0pt,bottom=0pt,colback=white,after=\ignorespacesafterend\par\noindent]
\noindent\textbf{RQ2. How are AutoML topics distributed across different machine learning life cycle phases?}
We manually annotate 348 questions into one of six AutoML life cycle phases by taking statistically significant sample questions from each of the four topic categories. We find that Model training, i.e., the implementation phase is the most dominant with 28.7\% questions, followed by Data Preparation (24.7\% questions), Model Designing (18.7\% questions), Model Deployment \& Monitoring (12.9\% questions), Model Evaluation (9.2\% questions), and Requirement Analysis, i.e., scope definition (5.7\% questions).
We find that during the Implementation phase troubleshooting, an error message or stack traces-related questions dominate. However, during the Requirement Analysis, Model Design, and Model Deployment phases, most queries are related to different features or limitations of AutoML services. So better tutorials and learning resources may help the AutoML practitioners in the early stages of the ML Pipeline.
\end{tcolorbox}

\subsection{RQ3. What AutoML topics are most popular and difficult on SO?}\label{rq_diff_pop}
\subsubsection{Motivation}
Our previous analysis shows that AutoML practitioners face many challenges related to machine learning development and specific challenges to AutoML tools/platforms (e.g., MLOps, Model development, and debugging). Some of these posts are more common than others, i.e., more community participation (i.e., view counts, up-votes, etc.). So, challenges faced for a particular topic during a specific MLLC phase are not equally difficult to get a working solution. A thorough analysis of the popularity and difficulty of practitioners' challenges may yield valuable insight. This analysis may help the researchers and AutoML providers to prioritize efforts to take necessary actions on the usability and applicability of these tools/platforms to make to more accessible to practitioners, especially domain experts.

\subsubsection{Approach}
In this research, we use the following five attributes from our SO dataset to determine the popularity or difficulty of a group of questions. 
We compute the difficulty for a group of questions using two attributes for each of the questions in that group \begin{inparaenum}[(1)] 
\item The percentage of questions that do not have an accepted answer,
\item Average median time to get a solution.
\end{inparaenum}
Similarly, we estimate how popular the topic is throughout the SO community using the following three popularity metrics: \begin{inparaenum}[(1)]
\item Average \#views, 
\item Average \#favorites, 
\item Average score. 
\end{inparaenum}

These five metrics are common attributes of a SO question, and numerous research in the field~\cite{iot21, alamin2021empirical, bagherzadeh2019going, abdellatif2020challenges, ahmed2018concurrency} have utilized them to assess a question's popularity and the difficulty of finding a solution. One question may have several answers on SO, and the user who asked the question can mark one reply as correct. As a result, the accepted response is regarded as accurate or of high quality. Therefore, the lack of an accepted response can mean the user could not find a suitable, useful response. The proper problem description, i.e., the quality of the question, helps to get an accepted answer. However, the main reason for not having an accepted answer is most likely because the SO community find that problem, i.e., the question difficult. A crowd-sourced platform like SO depends on its users' ability to supply accurate, usable information swiftly. In SO, the median time to receive an answer is only about 21 minutes~\cite{iot21}, but questions from a challenging or domain-specific topic may require more time to be answered.

It might be challenging to evaluate the topics' popularity and the difficulty of obtaining an accepted answer using a variety of measures. As a result, following related study~\cite{iot21}, we calculate two fused metrics for topic popularity and difficulty. The two fused metrics are described below.

\bf{Fused Popularity Metrics.} We begin by calculating the popularity metrics for each of the thirteen AutoML topics. However, the range of these attributes varies a lot. For example, the average number of views is in the range of hundreds, average scores may range from zero to five, and the average number of favorites might range from zero to three. Therefore, following related study~\cite{iot21}, we normalize the values of the metrics by dividing them by the mean metric value across all groups (e.g., for topics $K$ = 13). In this way, we get their normalized popularity metrics: $View\_N_{i}$, $Favorite\_N_{i}$, and $Score\_N_{i}$ for each topic $i$ (e.g., AutoML topics $K$=13). Finally, following related work~\cite{iot21}, we calculate the average fused popularity metric $Fused\_Popularity{_i}$ for each topic.
\begin{eqnarray}
View\_N_{i} = \frac{View_{i}}{\frac{\sum_{j=1}^{K}View_j} {K}} \\
Favorite\_N_{i} = \frac{Favorite_{i}}{\frac{\sum_{j=1}^{K}Favorite_j} {K}} \\
Score\_N_{i} = \frac{Score_{i}}{\frac{\sum_{j=1}^{K}Score_j} {K}}
\end{eqnarray}
\begin{equation}\label{eq:fusedP}
Fused\_Popularity{_i} = \frac{View\_N_{i} + Favorite\_N_{i} + Score\_N_{i}}{3}
\end{equation}

Similar to popularity measurements, we begin by calculating the complexity metrics for each issue. Then, the metric values are normalized by dividing them by the group-wide average metric value (e.g., 40 for LCSD topics). Consequently, we generate two new normalized measures for a given topic $i$. Finally, the fused difficulty metric $FusedD i$ of topic $i$ is calculated by averaging the normalized metric values.

\bf{Fused Difficulty Metrics.} Similar to topic popularity metrics, we begin by calculating the difficulty metrics for each topic. Then, we compute the normalized difficulty metrics for a given topic $i$ by dividing them by the mean value across all groups (e.g., 13 topics). Finally, We compute the fused difficulty metric $Fused\_difficulty_{i}$ of topic $i$ by calculating the average of these two normalized metrics.

\begin{eqnarray}
PctQuesWOAccAns\_N_{i} = \frac{PctQWoAcceptedAnswer_{i}}{\frac{\sum_{j=1}^{K}PctQWoAcceptedAnswer_{j}}{K}} \\
MedHrsToGetAccAns\_N_{i} = \frac{MedHrsToGetAccAns_{i}}{\frac{\sum_{j=1}^{K}MedHrsToGetAccAns_j}{K}}
\end{eqnarray}   
\begin{equation}\label{eq:fusedD}
Fused\_difficulty_{i} = \frac{PctQuesWOAccAns\_N_{i} + MedHrsToGetAccAns\_N_{i}}{2}
\end{equation}

Additionally, we aim to establish a relationship between the difficulty and popularity of these topics. In this study, we utilize the Kendall Tau correlation~\cite{Kendall-TauMetric-Biometrica1938} to determine the relationship between the popularity and difficulty of a topic. In contrast to the Mann-Whitney correlation~\cite{kruskal1957historical}, it is not subject to outliers in the data. We can not analyze the popularity and difficulty of these topics because SO does not provide times data for these popularity metrics~\cite {iot21, alamin2021empirical, abdellatif2020challenges}.

\subsubsection{Result}\label{rq_result:pop_diff}

\begin{figure}[t]
\centering
\includegraphics[scale=0.60]{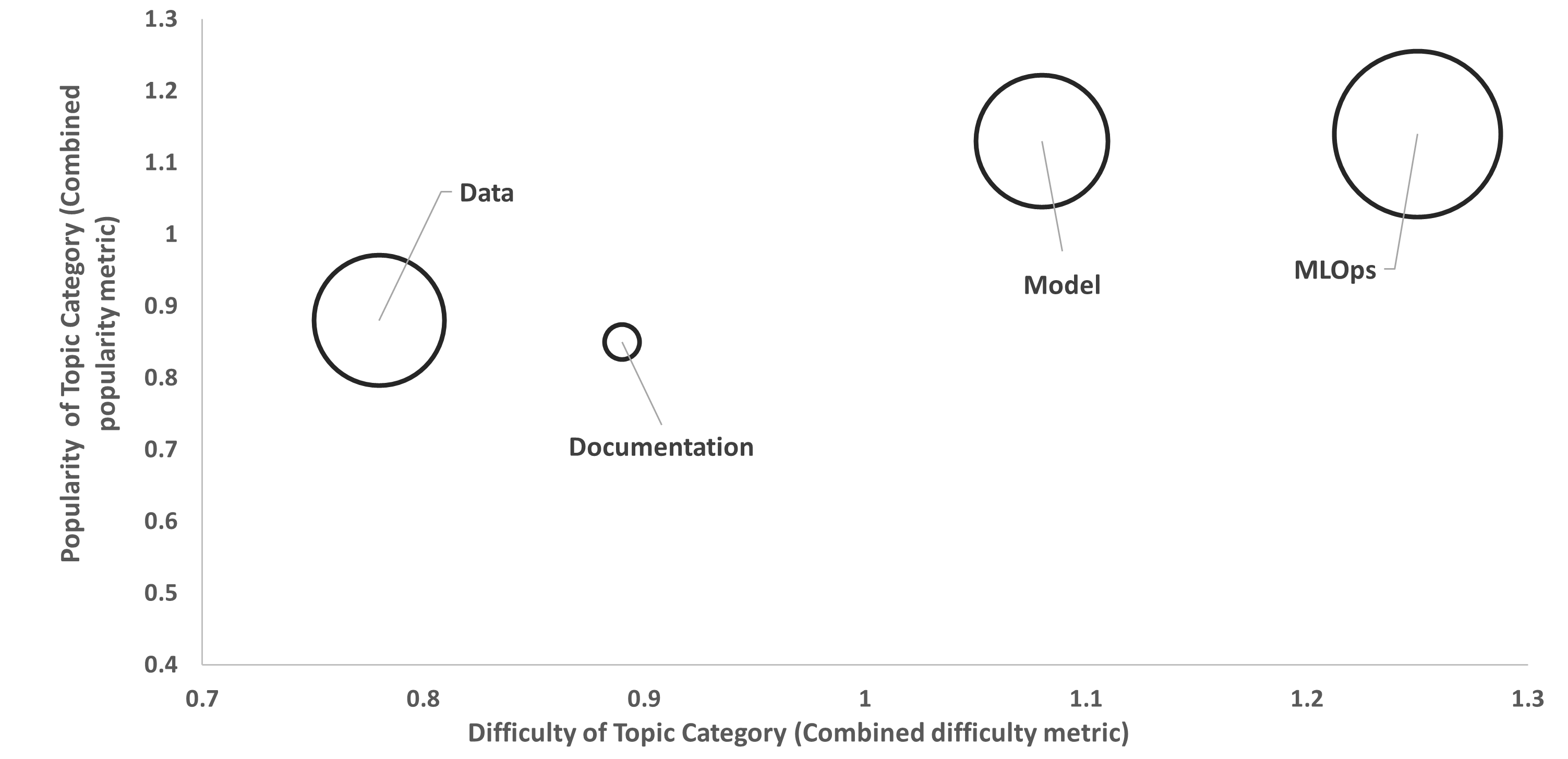}
\caption{The popularity vs. difficulty of AutoML topic categories.}
\label{fig:bubble_diff_pop_per_category}
\end{figure}

\begin{table}[t]
  \centering
   \caption{Popularity metrics for AutoML topics.}
    \begin{tabular}{llr|rrr}\toprule
    \textbf{Topic} & \textbf{Category} & {\textbf{Fused\_Popularity}} & {\textbf{\#View}} & {\textbf{\#Favorite}} & {\textbf{\#Score}} \\
    \midrule
        Model Performance & Model & 4.48 & 892.4 & 0.3 & 1.2 \\
        Model Load \& Deployment & MLOps & 4.05 & 665.7 & 0.3 & 1.1  \\ 
        Library/Platform Management & MLOps & 3.89 & 983.5 & 0.2 & 1.1  \\ 
        Data Transformation & Data & 3.87 & 1054.9 & 0.2 & 1 \\ 
        Resource Management & MLOps & 3.26 & 625 & 0.2 & 1 \\ 
        Pipeline Automation & MLOps & 3.11 & 513.9 & 0.2 & 1  \\ 
        Model Design & Model & 2.8 & 546.9 & 0.2 & 0.7  \\ 
        Model Debugging & Model & 2.77 & 786.7 & 0.1 & 0.9   \\ 
        Data Management & Data & 2.67 & 1047.8 & 0.1 & 0.5  \\ 
        Library Documentation &Documentation &2.61 &672.2 &0.1 &0.9 \\
        Bot Development &MLOps &2.51 &602.5 &0.1 &0.9 \\
        Data Exploration &Data &2.02 &671.4 &0.1 &0.4 \\
        Model Training \& Monitoring &Model &1.96 &376.9 &0.1 &0.7 \\

    \bottomrule
    \end{tabular}%
   \label{tab:topicPopularity}%
\end{table}%

\begin{table}[ht]
  \centering
   \caption{Difficulty for getting an accepted answer for AutoML topics}
    \begin{tabular}{llr|rr}\toprule
    \textbf{Topic} & \textbf{Category} & {\textbf{Fused\_difficulty}} & {\textbf{Med. Hrs To Acc.}} & {\textbf{Ques. W/O Acc.}} \\
    \midrule

    Model Load \& Deployment &MLOps &1.75 &45.9 &70 \\
    Model Training \& Monitoring &Model &1.4 &32 &72 \\
    Pipeline Automation &MLOps &1.09 &22.9 &63 \\
    Model Design &Model &1.05 &21.3 &63 \\
    Bot Development &MLOps &1.04 &19.2 &69 \\
    Resource Management &MLOps &0.97 &17.4 &65 \\
    Data Transformation &Data &0.96 &17.9 &63 \\
    Library/Platform Management &MLOps &0.9 &15.9 &62 \\
    Model Debugging &Model &0.88 &12.6 &70 \\
    Model Performance &Model &0.79 &12.1 &60 \\
    Data Exploration &Data &0.77 &11.6 &60 \\
    Library Documentation &Documentation &0.76 &14.2 &49 \\
    Data Management &Data &0.63 &5 &63 \\

    \bottomrule
    \end{tabular}%
  \label{tab:topicDifficulty}%
\end{table}%

\begin{table}[h]
\centering
\caption{Correlation between the topic popularity and difficulty.}
\begin{tabular}{lrrr}\toprule
coefficient/p-value & \bf{View} & \bf{Favorites} & \bf{Score}\\ \midrule
\bf{\% Ques. w/o acc. ans} &   -0.24/0.26 & -0.08/0.73 & -0.07/0.75 \\
\bf{Med. Hrs to acc. ans} &    -0.51/0.01 & 0.28/0.23 & 0.14/0.53  \\
\bottomrule
\end{tabular}%
\label{tab:pop_diff_correlation}%
\end{table}

In Figure~\ref{fig:bubble_diff_pop_per_category}, we present an overview of the popularity vs. difficulty of our four AutoML topic categories. 
The bubble size represents the number of questions, i.e., the bigger the bubble, the more questions the topic category has. 
It shows that MLOps is the most difficult topic category, followed by Models, Documentation, and Data. Similarly, MLOps is the most popular topic category, followed by Model, Data, and Documentation. Our observation suggests that AutoML practitioners find various aspects of developing ML models via AutoML tools/platforms popular as well as difficult related to improving the performance of the ML model (e.g., \dq{37196368}), deploying the trained model via API endpoint (e.g., \dq{50334563}). Similarly, we find that questions related to AutoML SO community have strong support for finding relevant API parameters via official documentation (e.g., \dq{60667920}) but relatively less support on the operational and management of these models.

\bf{Topic Popularity.} 
Table \ref{tab:topicPopularity} presents the three popularity metrics: Average number of 1. Views, 2. Favorites, 3. Scores. It also contains the combined popularity metrics (i.e., Fused\_Popularity) using Equation \ref{eq:fusedP}. In the Table, the AutoML topics are presented in descending order based on the Fused\_Popularity metric.
Model Performance topic from the Model topic category has the highest Fused\_Popularity score(i.e., 4.9), highest average favorite count (i.e., 0.3), and average score (i.e., 1.2). This is the third largest topic (e.g., 12.5\% of questions), and it reflects the common struggle of AutoML practitioners to enhance the model's performance using the best AutoML configuration (e.g., in \dq{59219818}). 
The Data Transformation topic under the Data category has the highest average view count (i.e., 1055), followed by Data Management (i.e., 1028) and Library/Platform Management (i.e., 984) from MLOps. Data filtering and prepossessing are crucial for optimal model performance, and our analysis reveals that AutoML practitioners regularly face these challenges. For example, in \dq{38927230}, a practitioner is struggling to text pre-processing using the Pandas library on the Azure cloud platform, which has around 45K views.
Model Training \& Monitoring from the Model category has the lowest Fused\_Popularity score(i.e., 1.96) and average favorite count (i.e., 0.1). This topic contains questions that are tool/platform-specific (e.g., \dq{51878538}, \dq{58177571}), and so attract a small number of AutoML practitioners.

\bf{Topic Difficulty.}
In Table~\ref{tab:topicDifficulty}, we present the summary of two difficulty metrics: 1. Percentage of questions without accepted \rev{answers}, 2. Median hours to get accepted answers for our 13 AutoML topics. Similar to topic popularity, we also report the combined topic difficulty metrics (e.g., Fused\_difficulty) using Equation \ref{eq:fusedD}. The AutoML topics in Table \ref{tab:topicDifficulty} are organized in descending order based on the Fused\_difficulty value.
Topic Model Load \& Deployment is the most difficult topic in terms of Fused\_difficulty metric (i.e., 1.75), median hours to get accepted answers (i.e., 45.9 hours), the second highest number of questions without accepted answers (i.e., 70\%). Model Training \& Monitoring topic from the Model category has the highest number of questions without any accepted answers (i.e., 72\%) followed by Model Load \& Deployment from MLOps category with (i.e., 70\%) and Model debugging (i.e., 70\%). These questions receive less help from the SO community since they are highly specific to a particular tool or platform (e.g., \dq{58320938}) and are difficult to reproduce for others. Thus they frequently lack accepted answers (e.g., \dq{65921153}, \dq{70950914}).
Library Documentation has the second least difficult topic in terms of Fused\_difficulty (i.e., 0.76) with the least amount of questions without accepted answers (i.e., 49\%). Frequently, AutoML practitioners lack the abilities required to discover relevant documentation, and the AutoML SO community is quite supportive in providing the required resources (e.g., \dq{40227519}, \dq{42446046}).
Data Management under the Data category is the least difficult topic with a Fused\_difficulty value of 0.63, the least median hours to get accepted answers (i.e., 5 hours). These queries are typical for data analytics (e.g., \dq{54896417}) and some pre-processing data tasks similar to traditional ML application development (e.g., \dq{24700708}), therefore a border ML community combined with AutoML practitioners can contribute accepted answers.

\bf{Correlation between topic Popularity vs. Difficulty.}
We systematically explore if there is any positive or negative co-relationship between topic popularity and difficulty. For example, we find that Model Load \& Deployment topic is the most difficult and the second most popular topic. Alternatively, we find that Model Training \& Monitoring is the least popular but the second most difficult topic.

Table~\ref{tab:pop_diff_correlation} presents the six correlation measures between topic fused difficulty and popularity metric presented in Table~\ref{tab:topicPopularity} and~\ref{tab:topicDifficulty}. Four out of six correlation coefficients are negative, and the other two are positive; thus, they are not statistically significant, with a 95\% confidence level. Therefore, from this analysis, we can not say the most popular topic is the least difficult to get an accepted answer to and vice versa. Nonetheless, AutoML researchers and platform providers could use this insight to take further necessary steps to accept solutions to most of the popular AutoML topics.

\begin{tcolorbox}[flushleft upper,boxrule=1pt,arc=0pt,left=0pt,right=0pt,top=0pt,bottom=0pt,colback=white,after=\ignorespacesafterend\par\noindent]
\noindent\textbf{RQ3. What AutoML topics are most popular and difficult on SO?}
MLOps is the most challenging category, followed by Model, Documentation, and Data. We find that model performance improvement-related topics and operational aspects of ML pipeline, i.e., the configuration of library and packages and model deployment-related topics, are most common and popular among AutoML practitioners. Similarly, we also find that AutoML practitioners find Automated ML pipeline, model deployment \& monitoring related topics most difficult. In summary, we find that the topics that are more specific to AutoML services, i.e., have relatively greater difficulty than those that are generic to all tools/platforms or ML development.
\end{tcolorbox}
\subsection{RQ4. How does the topic distribution between cloud and non-cloud AutoML services differ?}\label{rq:tool_vs_platform}
\subsubsection{Motivation}
Both cloud-based and non-cloud-based AutoML services are present in our dataset. Each service provides a unique collection of services; their use cases are distinct. Cloud-based solutions, for instance, emphasize an end-to-end pipeline for data collecting to model operationalization, whereas non-cloud-based services offer greater customization for high-level data preparation, feature engineering, model evaluation, and model updating.
In this study, we investigate how AutoML-related topics, MLLC life cycle phases, and topic popularity and complexity vary between cloud-based and non-cloud-based AutoML solutions.
This analysis will provide valuable insights into the challenges of cloud versus non-cloud low-code (i.e., AutoML) services and opportunities for further development of these solutions.

\subsubsection{Approach}
In order to understand how the challenges of cloud vs. non-cloud AutoML solutions differ, first, we need to separate cloud vs. non-cloud AutoML SO questions. So our approach is divided into two main steps: \begin{inparaenum}[(1)]
        \item Step 1. Identify SO tags for cloud vs. non-cloud solutions,
        \item Step 2. Separating SO questions,
    \end{inparaenum}

\nd \textbf{Step 1. SO tags separation.}
In order to accomplish this, we manually analyze the 41 AutoML-related tags, as explained in Section~\ref{sec:methodology}, to determine which SO tags correspond to cloud-based or non-cloud-based AutoML solutions. For instance, we classify the ``amazon-sagemaker'' SO tag as a cloud solution because it contains discussions about the Amazon cloud SageMaker solution, whereas the ``pycaret'' SO tag is classified as a non-cloud solution because it contains discussions about the python-based, non-cloud AutoML framework PyCaret.

\nd \textbf{Step 2. SO questions separation.}
We must separate SO questions; however, a SO question may contain between one to five SO tags. We consider a query to belong to the non-cloud category if it has both cloud and non-cloud tags, as this provides us with greater granularity. For example, (e.g., ``How can I automate Splunk iterations using REST API'' in \dq{48421583}) contains SO tags ``splunk'', ``splunk-sdk'' which cloud and non-cloud tags respectively. Considering these queries as non-cloud (i.e., AutoML tools/framework) serves our objective more effectively.

After this, we analyze how AutoML topic categories, MLLC phases, and topic popularity and difficulty differ in the cloud vs. non-cloud solutions. For cloud vs. non-cloud distribution of MLLC phases, we use the same 348 annotated SO questions described in Section~\ref{sub-sec:mllc}.

\subsubsection{Result}

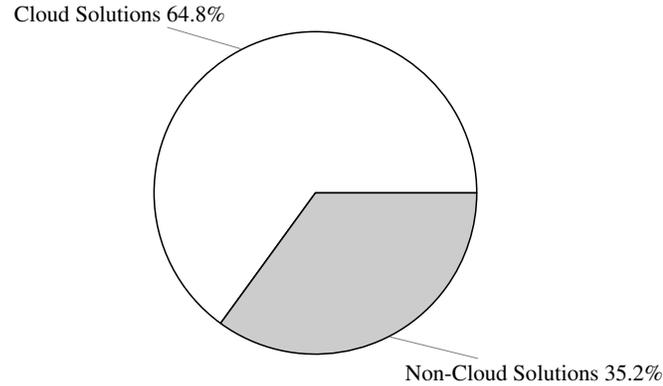
\begin{figure}[t]
	\centering
	\resizebox{3.5in}{!}{
	\begin{tikzpicture}
    \pie[
        explode=0.0, text=pin, number in legend, sum = auto, 
        color={black!0, black!20},
        ]
        {            
            65/\LARGE{Cloud Solutions} 64.8\%,
            35/\LARGE{Non-Cloud Solutions} 35.2\%
        }
    \end{tikzpicture}
    }
	\caption{Distribution of questions on cloud vs non-cloud AutoML services.}
	\label{fig:distribution_of_cloud_vs_noncloud}
\end{figure}

\begin{figure}[t]
\centering
\includegraphics[scale=.6]{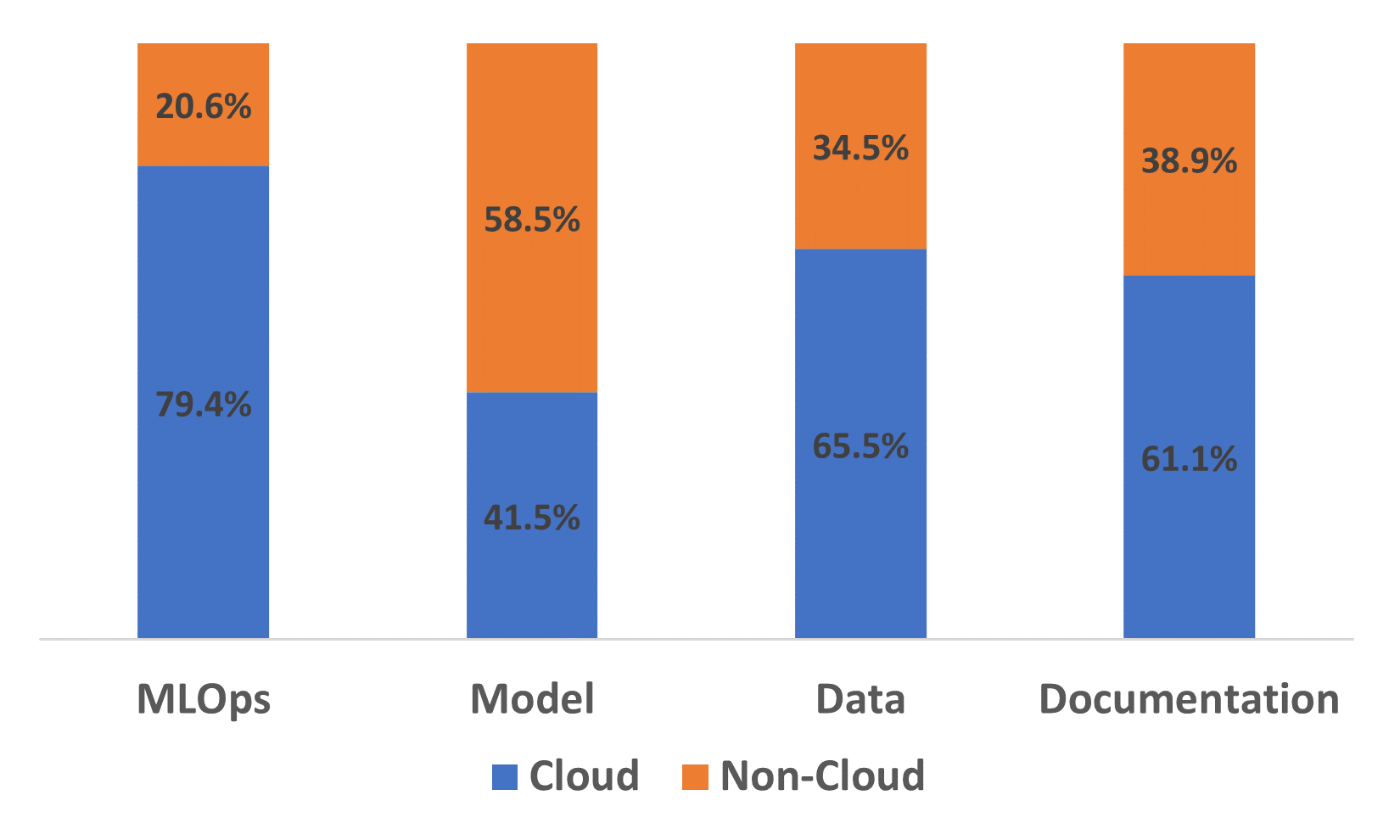}
\caption{ The distribution of AutoML topic categories between cloud vs non-cloud AutoML services. }
\label{fig:cat_cloud_vs_noncloud}
\end{figure}

\begin{table}[t]
  \centering
   \caption{Distribution (frequency) of MLLC phase between cloud vs non-cloud AutoML services. Each colored bar denotes a phase 
   (Black = Cloud, Green = Non-cloud Solution)}
    \begin{tabular}{lr}
    \toprule{}
    \textbf{MLLC Phases} & \textbf{\#Questions in cloud vs non-cloud solution}\\
    \midrule
        \textbf{Requirement Analysis} & \twobars{13}{7} \\
        \textbf{Data Preparation} & \twobars{60}{27} \\
        \textbf{Model Designing(MD)} & \twobars{36}{29} \\
        \textbf{Model Training (MT)} & \twobars{54}{45} \\
        \textbf{Model Evaluation (ME)} & \twobars{12}{20} \\
        \textbf{Model Deployment and Monitoring (MDM)} & \twobars{36}{9} \\
    \bottomrule
    \end{tabular}%
  \label{tab:topicMLLC_cloud_noncloud}%
\end{table}%

\begin{figure}[htbp]
\subfloat[A comparison of difficulty between cloud vs non-cloud topic categories]{\includegraphics[height=2.5in, width=2.5in]{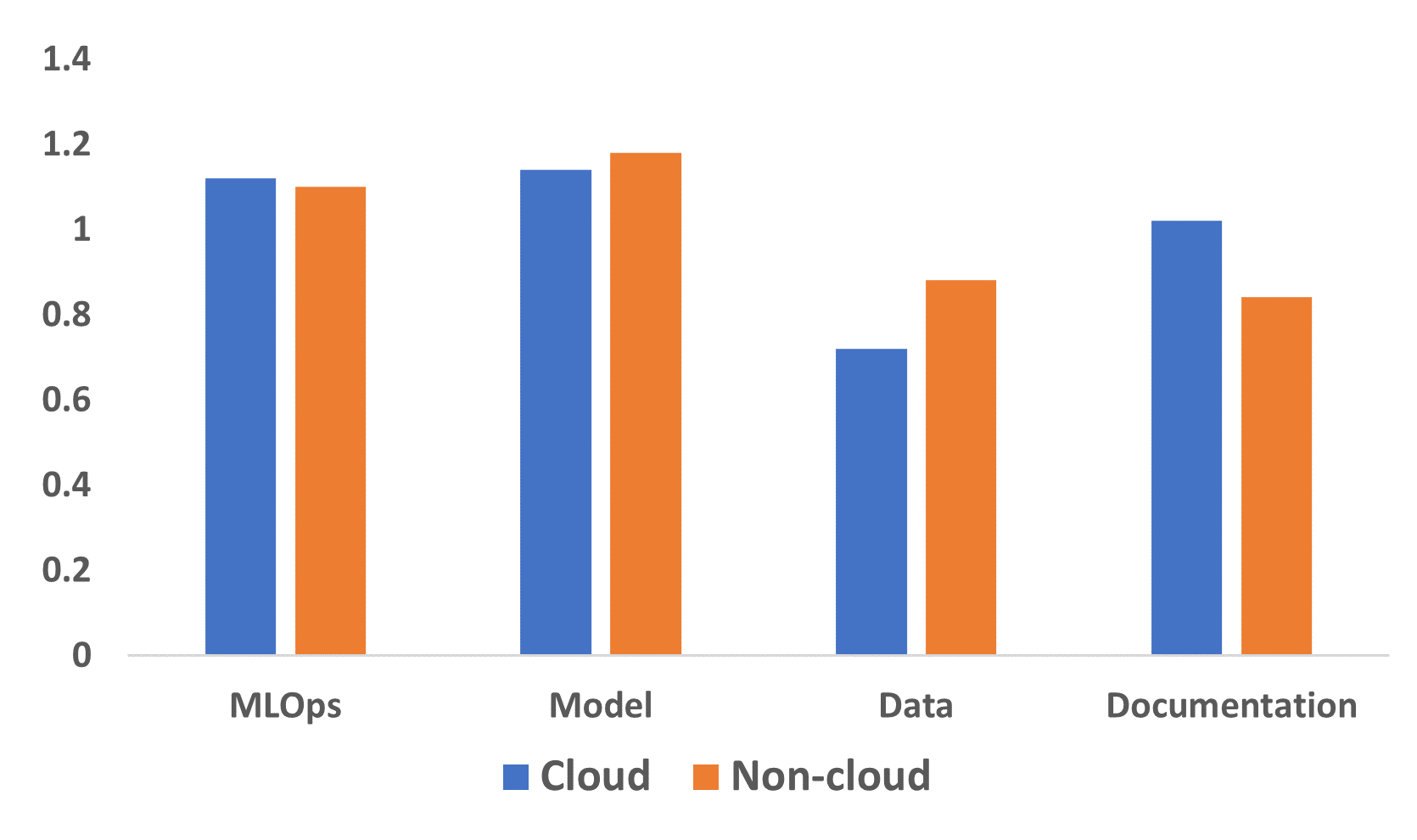}}
\qquad\qquad
\subfloat[A comparison of popularity between cloud vs non-cloud topic categories]{\includegraphics[height=2.5in, width=2.5in]{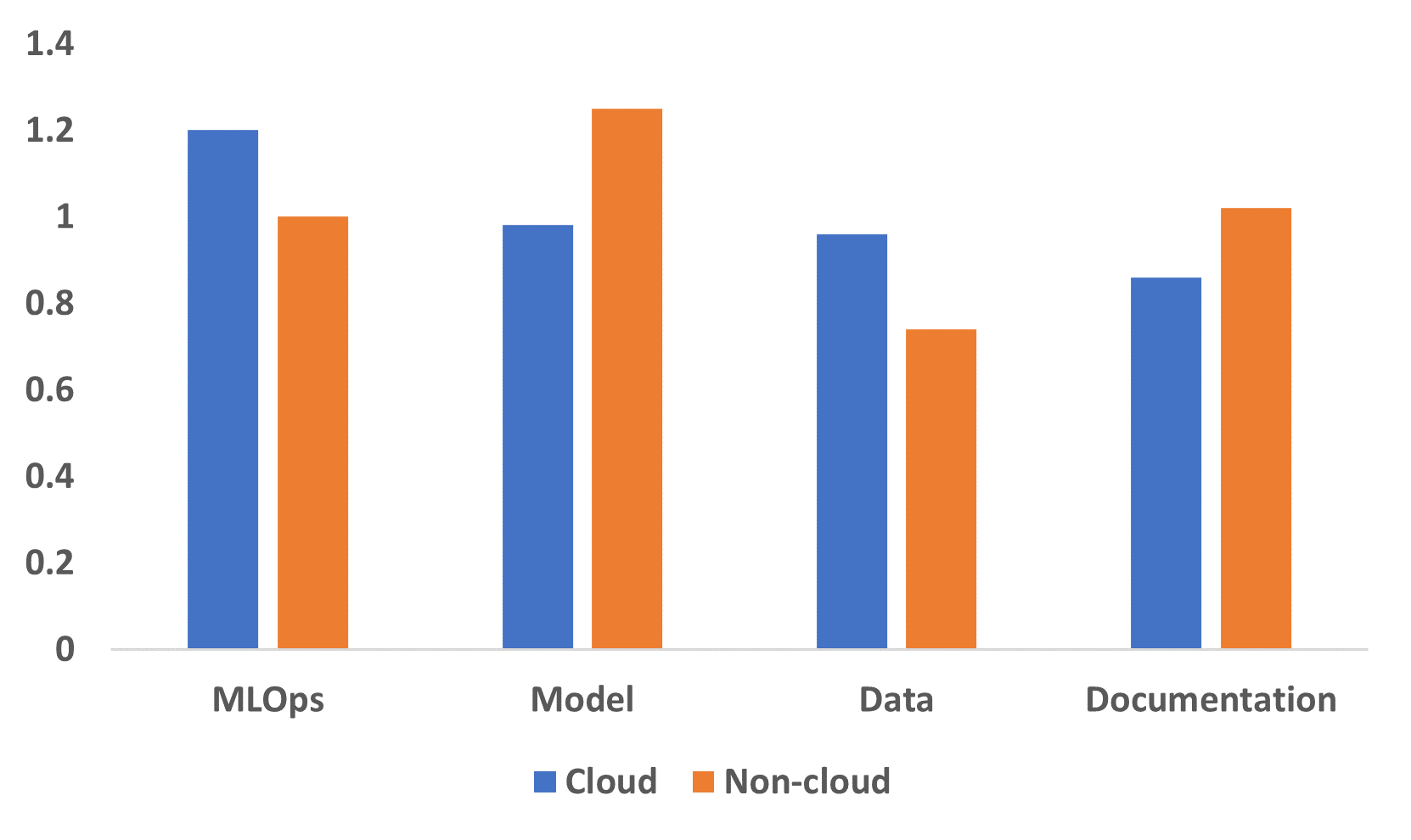}}
\caption{An overview of combined difficulty and popularity between cloud vs non-cloud AutoML solution's topic categories.}
\label{fig:cloud_vs_noncloud_pop_diff}
\end{figure}

Figure~\ref{fig:distribution_of_cloud_vs_noncloud} offers an overview of the cloud versus non-cloud AutoML service SO question in our dataset. After manual analysis, we find 23 SO tags out of 41 are related to cloud-based solutions, while the remaining 18 are related to non-cloud-based solutions. We observe that 6,833 (i.e., 64.8\%) SO questions belong to the cloud, and 3,716 (i.e., 35.2\%) SO questions belong to non-cloud AutoML solutions. This insight demonstrates the prevalence and popularity of cloud-based AutoML services.

Figure~\ref{fig:cat_cloud_vs_noncloud} shows the distribution of AutoML topic categories between cloud vs. non-cloud services. We can see that for the MLOps category, most of the questions (i.e., 80\%) belong to cloud platforms. Many of these questions belong to model deployment (e.g., ``How can I deploy an ML model from ECS to Sagemaker?'' in \dq{48537811}) and  (e.g., ``Monitoring the performance of ML model on EC2 Instance'' in \dq{66655967}).
Similarly, we can see that for the Data category, cloud-based solutions dominate (66\%) with queries related to cloud data management (e.g., ``Load Azure ML experiment run information from datastore'' in \dq{66801546}).
Non-cloud-based services dominate (59\%) on Model related queries (e.g., ``Retraining a model in PyCaret'' in \dq{68330216}), (e.g., ``Autokeras training model 10x slower (Google Colab)'' in \dq{68878481}). AutoML cloud solutions also have similar queries related to the Model's training performance (e.g., ``Training on Sagemaker GPU is too slow'' in \dq{67213383}).
For the Documentation category, around 62\% of questions belong to cloud-based AutoML services where practitioners ask queries like (e.g., ``API call for deploying and undeploying Google AutoML natural language model - documentation error?'' in \dq{67470364}). Non-cloud-based services practitioners ask questions requesting detailed documentation in \dq{60667920}.

Table~\ref{tab:topicMLLC_cloud_noncloud} demonstrate the distribution of cloud vs. non-cloud solution across different MLLC phases. Among these 348 SO questions, 211 (i.e., 60\%) questions belong to AutoML cloud solutions and 137 (i.e., 40\%) questions belong to non-cloud AutoML solutions.
We observe that practitioners using AutoML cloud solutions dominate queries during the Model Deployment and Monitoring (82\%) Phase, (e.g., ``how to deploy the custom model in amazon sageMaker'' in \dq{61248115}, deployment failure issue in \dq{61683506}).
We also observe that during the Model Evaluation phase, non-cloud AutoML solution-related queries dominate (63\%) (e.g., ``[H2O] How do I evaluate a multinomial classification model in R?'' in \dq{43556802}, ``Evaluate AutoKeras model gives different results then the same model as pure Keras evaluate (h5 file)'' in \dq{54491965}).
During other MLLC phases, the proportion of questions about cloud versus non-cloud solutions remains almost similar to their original distribution (e.g., Requirement Analysis (65\% vs. 35\%), Data Preparation (57\% vs. 43\%) and Model Training phase (55\% vs. 45\%), Model Designing (MD) phase (55\% vs. 45\%).

Figure~\ref{fig:cloud_vs_noncloud_pop_diff}  presents the comparison of popularity and difficulty (Section~\ref{rq_diff_pop}) of topic categories of AutoML cloud vs. non-cloud solutions. For the Documentation subject category AutoML, cloud-related inquiries are slightly more challenging (fused difficulty metric 1.02 vs. 0.84); however, for the Data topic category AutoML, non-cloud-related queries are more challenging (i.e., fused difficulty metric 0.88 vs. 0.72). For MLOps and Model, cloud and non-cloud-related queries are almost equally challenging.
According to the fused popularity metric, AutoML cloud solution-related queries on MLOps (i.e., 1.2 vs. 1.0 ) and Data (i.e., 0.96 vs. 0.74) topic categories are somewhat more popular. In contrast, AutoML non-cloud related SO queries on Model (i.e., 1.25 vs. 0.98) and Documentation (i.e., 1.02 vs. 0.86) topic categories are more popular.

\begin{tcolorbox}[flushleft upper,boxrule=1pt,arc=0pt,left=0pt,right=0pt,top=0pt,bottom=0pt,colback=white,after=\ignorespacesafterend\par\noindent]
\noindent\textbf{RQ4. How does the topic distribution between cloud and non-cloud AutoML services differ?}
In our dataset, around 64.8\% SO questions belong to the cloud, and 35.2\% belong to non-cloud-based AutoML solutions.
MLOps (i.e., 80\%), Data (i.e., 66\%), and Documentation (i.e., 61\%) topic categories are dominated by cloud-based AutoML solutions. They are comparable in terms of difficulty but marginally more popular, except for the Documentation category, which is marginally more challenging and less popular. Alternatively, the Model (59\%) topic category is predominant and popular across non-Cloud-based AutoML solutions.
For the Model Deployment and Monitoring phase, AutoML cloud solutions-related questions predominate (82\%). However, during the Model Evaluation phase, non-cloud AutoML solution-related queries predominate (63\%).
\end{tcolorbox}

\section{Discussions} \label{sec:discussion}

\begin{figure}[t]
\centering
\includegraphics[scale=0.55]{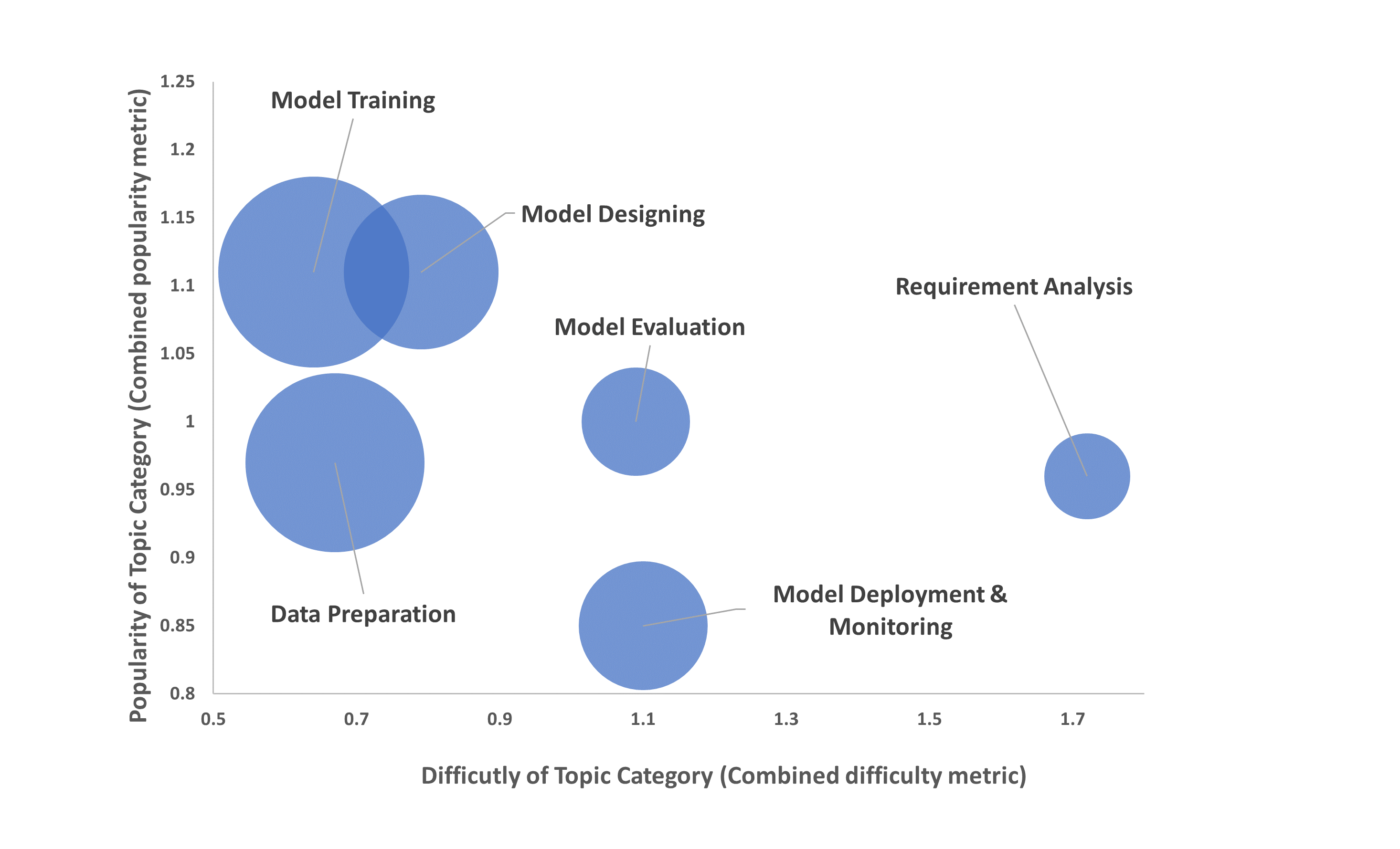}
\caption{The popularity vs. difficulty for AutoML pipeline phases.}
\label{fig:bubble_diff_pop_MLLC}
\end{figure}
\begin{figure}[t]
\centering
\includegraphics[scale=0.65]{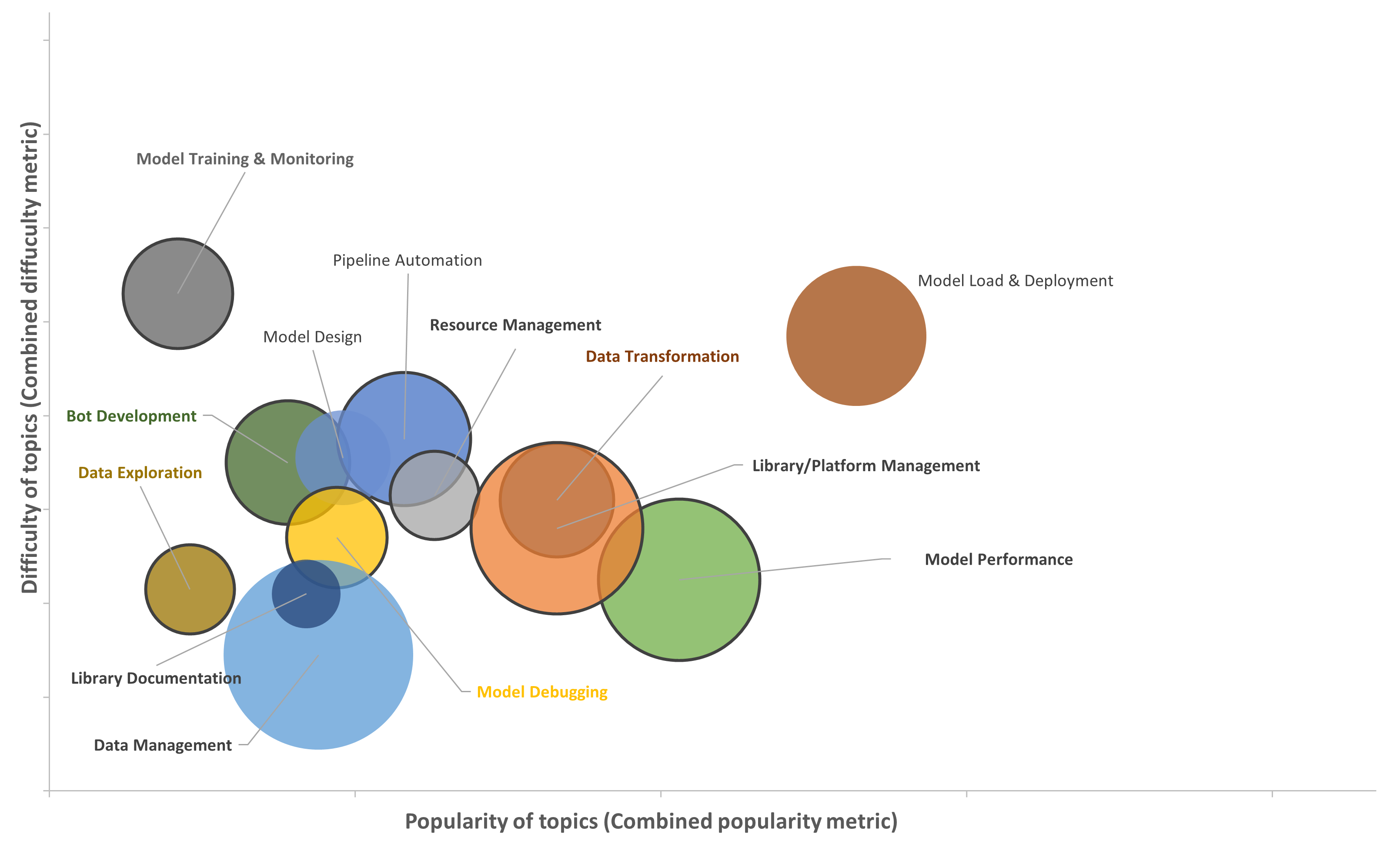}
\caption{The popularity vs. difficulty for AutoML topics.}
\label{fig:bubble_diff_pop_topic}
\end{figure}

\subsection{Evolution of AutoML related discussion}
\begin{figure}[t]
\centering
\includegraphics[scale=0.55]{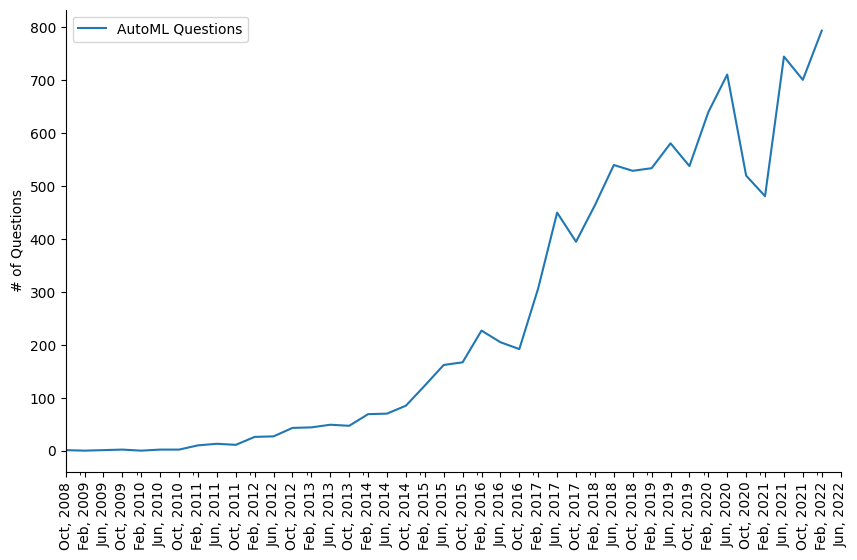}
\caption{The evolution of overall AutoML related questions over time.}
\label{fig:all_questions_evolution}
\end{figure}

\begin{figure}[t]
\centering
\includegraphics[scale=0.55]{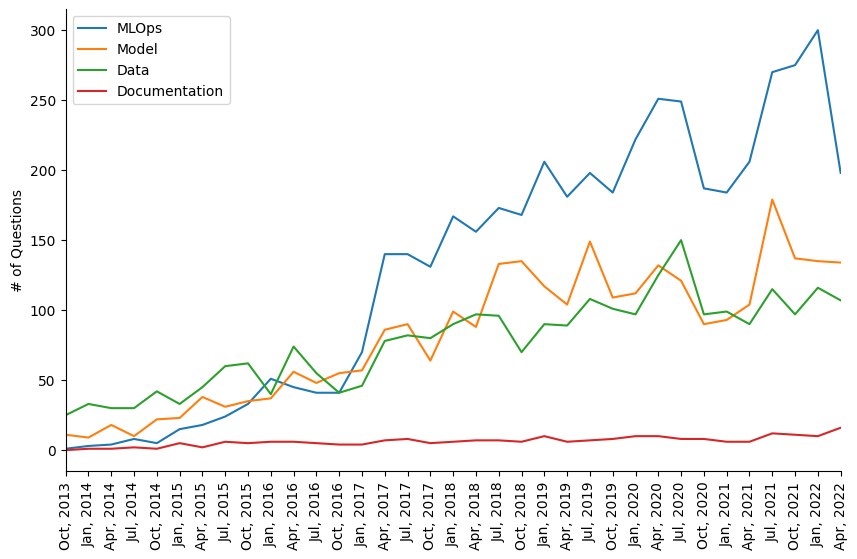}
\caption{The evolution of SO questions on AutoML topic categories over time.}
\label{fig:topic_cat_evolution}
\end{figure}

Figure~\ref{fig:all_questions_evolution} depicts the progression of overall  AutoML-related discussion from our extracted dataset between 2008 to 2022. Additionally, it demonstrates that AutoML-related discussion is gaining popularity in mid-2016 (i.e., more than 200 questions per quarter).
Figure~\ref{fig:all_questions_evolution} shows that during the first quarter of 2021, the total number of AutoML questions on Stack Overflow experiences a considerable decrease (i.e., about 25\%), primarily due to the low number of queries in Splunk and Amazon Sagemaker during the pandemic. However, Amazon launches a suite of new AutoML services at the start of 2021, and the overall trend for AutoML-related discussion shows an upward trend~\cite{aws_news}. We provide a more detailed explanation for Figure~\ref{fig:all_questions_evolution} below.

\textbf{MLOps}
This is the largest AutoML topic category, with around 43\% of AutoML-related questions and five AutoML topics. From Figure~\ref{fig:topic_cat_evolution}, we can see this topic category began to gain popularity from the start of 2017, i.e., after big tech companies started to offer AutoML cloud platform services (\sec\ref{sub-sec:automl}). From 2017 to 2022, AutoML library and cloud platform management-related topics remain the most popular in this category (e.g., issues with installing the library in different platforms \dq{61883115}).
From the beginning of 2017 to mid-2018 AutoML cloud platform's bot development-related queries were quite popular (e.g., ``Filling data in response cards with Amazon Lex'' in  \dq{47970307}). Around that time, Amazon Lex (2017) was released, a fully managed solution to design, build, test and deploy conversational AI solutions.
From our analysis, we can see the rise of Pipeline Automation-related discussion from the beginning of 2019 (e.g., issues related to the deployment pipeline in \dq{55353889}).
We also notice a significant rise in the number of questions related to model load and deployment from 2019 to 2022 (e.g., ``Custom Container deployment in vertex ai'' in \dq{69316032}).

\textbf{Model}
This is the second largest AutoML topic category, with around 28\% of AutoML-related questions and four AutoML topics.
From Figure~\ref{fig:topic_cat_evolution}, we can see that, similar to MLOps; the Model topic category began to gain popularity from the start of 2017. The dominant topic in this category is the Model performance improvement-related queries using AutoML services (e.g., ``Grid Search builds models with different parameters (h2o in R) yields identical performance - why?'' in \dq{47475848}, ``How to understand the metrics of H2OModelMetrics Object through h2o.performance'' in \dq{43699454}). From our analysis, we see around a 200\% increase in the number of questions in Model training and monitoring topic from 2021 to mid-2021. So practitioners ask queries regarding training issues (e.g., ``Training Google-Cloud-Automl Model on multiple datasets'' in \dq{68764644}, \dq{56048974})  from practitioners in recent times. Other topics related to model design, model implementation, and debugging evolved homogeneously around this time. 

\textbf{Data}
This is the third largest AutoML topic category, with around 27\% of AutoML-related questions and three AutoML topics. The Data Management topic is the most dominant topic in this category. From our analysis, we can see around a 60\% increase in the number of queries on cloud data management, especially after Splunk cloud becomes available on the google cloud platform. So, we can see developers' queries related to data transfer (e.g., in \dq{63151824}), REST API (e.g., in \dq{63152602}), API rate limitation (e.g., in \dq{63292162}) becomes popular. Other topics from this category evolve homogeneously over time.

\textbf{Documentation}
This is the smallest AutoML topic category with around 2\% of AutoML-related questions and one AutoML topic. We find that Since 2015 questions from this category shifted slightly from API-specific details (e.g., \dq{35562896}) to documentation-related limitations for cloud deployment (e.g., \dq{59328925}), containerization (e.g., \dq{68888281}).


\subsection{Top AutoML service providers} \label{sub_sec:top_automl}
\begin{figure}[]
\centering
\includegraphics[scale=0.55]{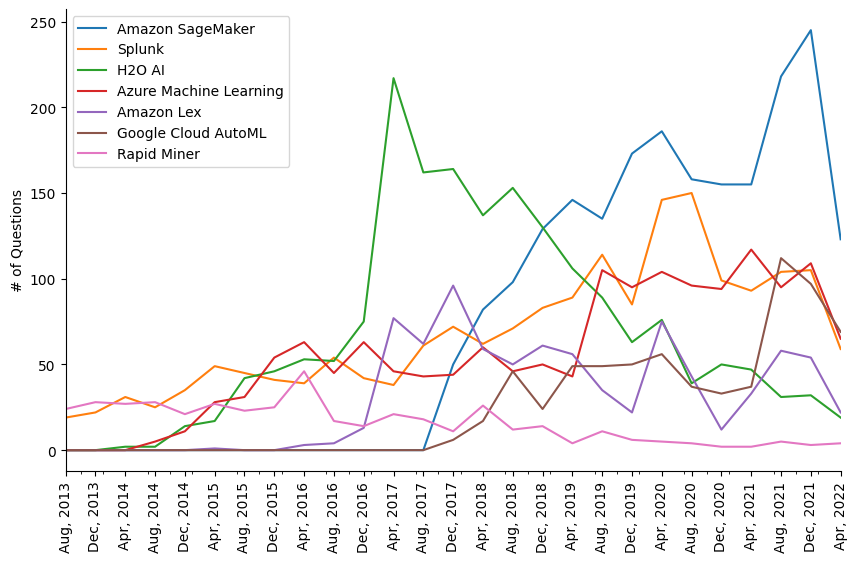}
\caption{The evolution of questions for top AutoML services.}
\label{fig:top_automl_platforms}
\end{figure}

In Figure~\ref{fig:top_automl_platforms}, we present the evolution of the top seven AutoML solutions in terms of the number of SO questions over the past decade. In our dataset, Amazon Sagemaker~\cite{aws_sagemaker}, is the largest AutoML platform containing around 20\% of all the SO questions followed by Splunk (19\%)~\cite{splunk}, H2O AI (17\%)~\cite{h2oai}, Azure Machine Learning~\cite{azureml}, Amazon Lex~\cite{aws_lex}, Google Cloud AutoML~\cite{google_automl}, and Rapid Miner~\cite{rapid_miner}. Among these AutoML service providers, H2O AI AI began in mid-2017 with the release of their H2O.ai driverless AI cloud platform, but since then, it has been steadily losing practitioners' engagement in SO. We also observe that from its release in 2017 AWS SageMaker cloud platform remains dominating, with a notable surge occurring in the middle of 2021. Other AutoML platforms likewise demonstrate an overall upward trend.

\bf{AutoML service providers' support on SO discussion.}
\begin{figure}[t]
\centering
\includegraphics[scale=0.45]{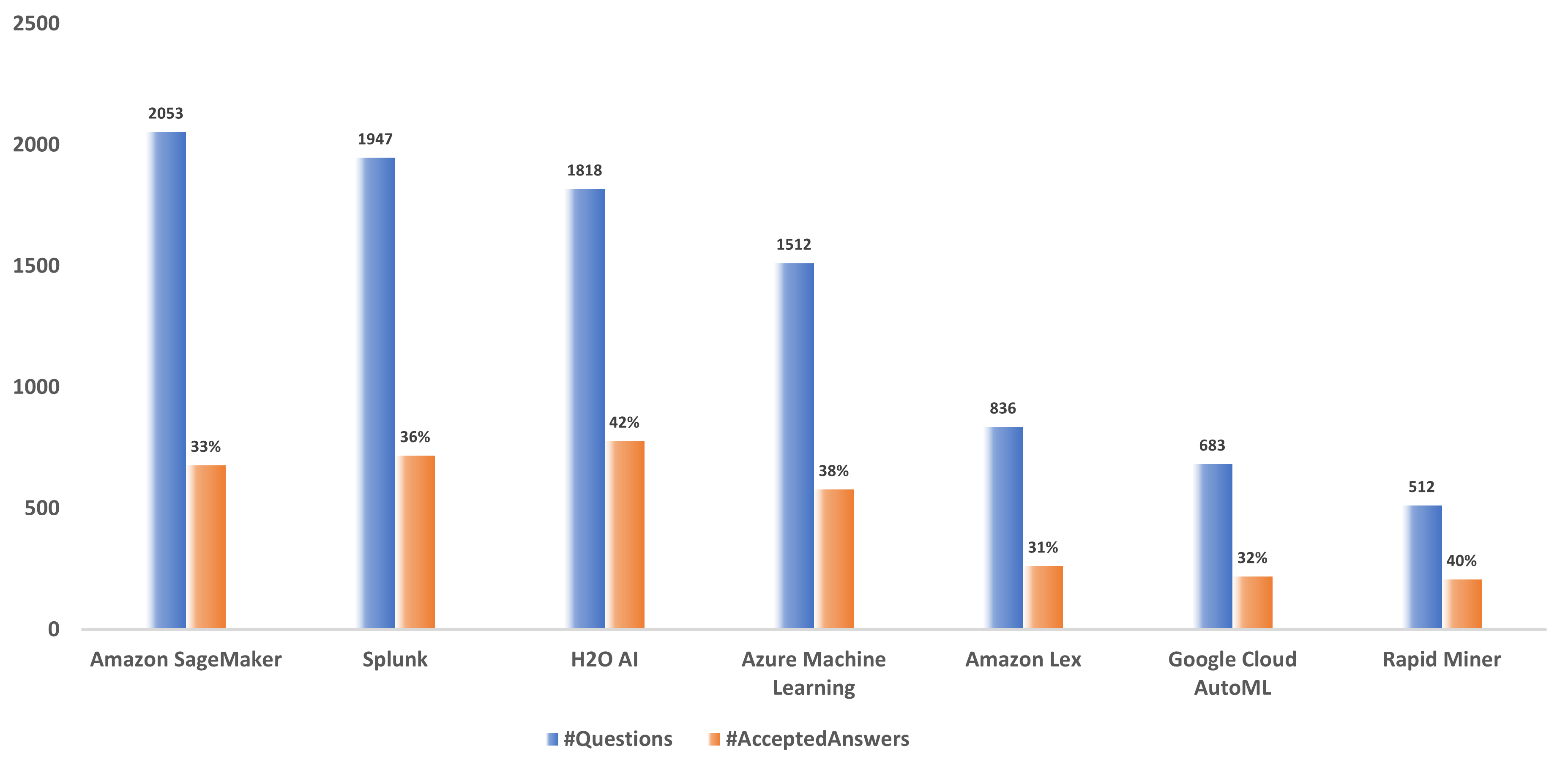}
\caption{The status of top platforms and their percentage of accepted answers.}
\label{fig:top_platform_acc_stat}
\end{figure}
\begin{figure}[t]
\centering
\includegraphics[scale=0.30]{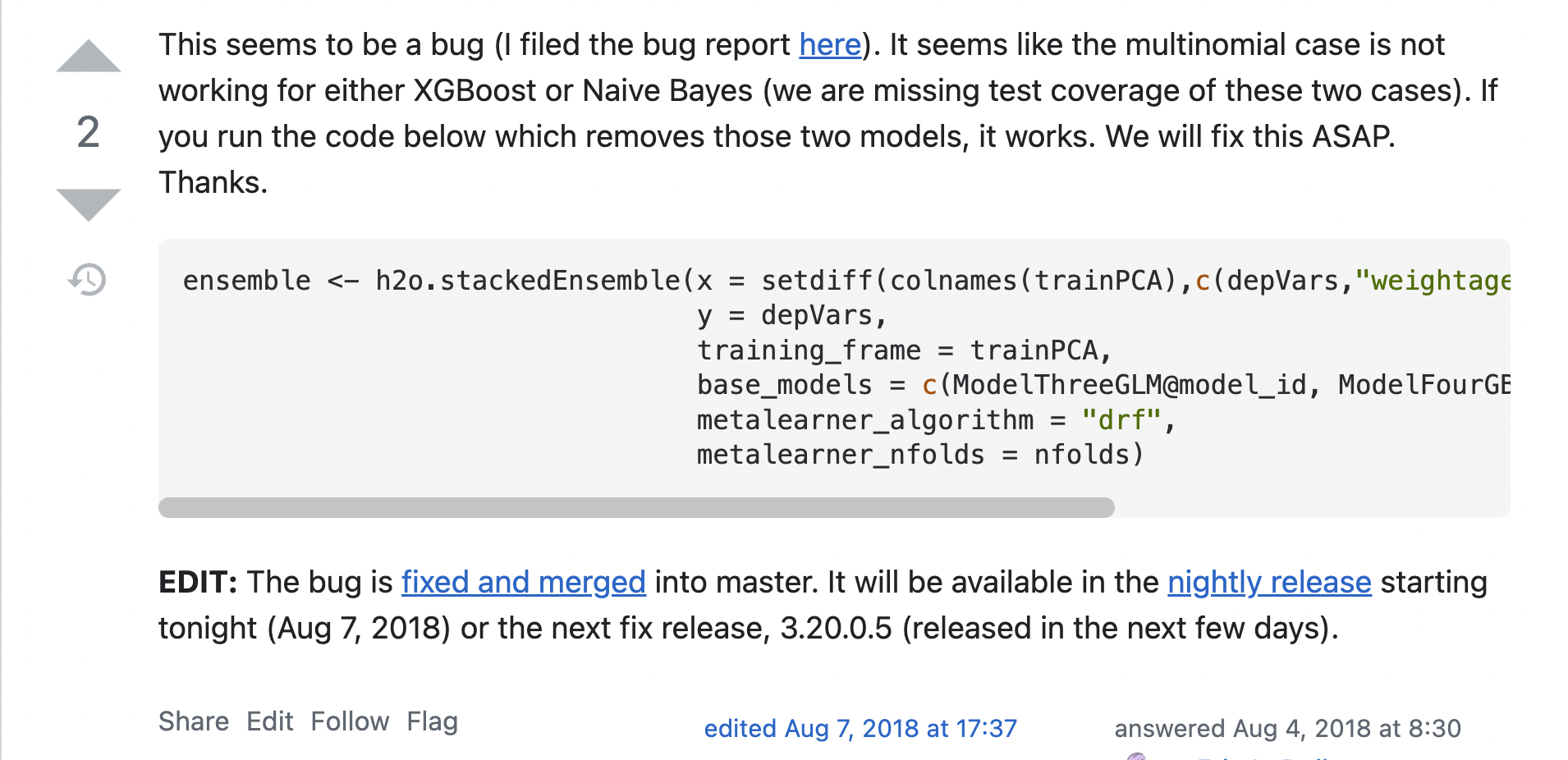}
\caption{An example of H2O's platform's support for resolving reported bugs (\da{51683851}).}
\label{fig:h2o_support}
\end{figure}
\begin{figure}[t]
\centering
\includegraphics[scale=0.30]{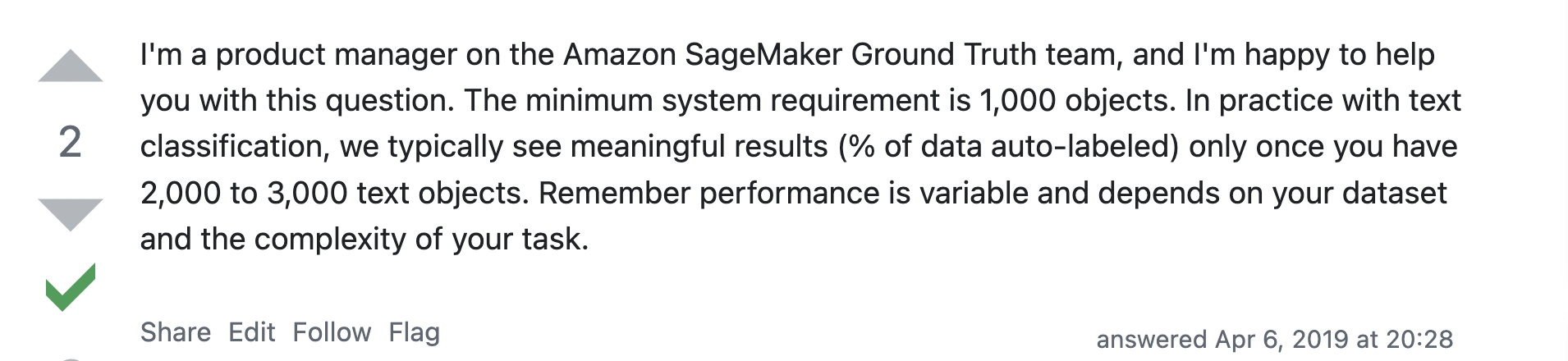}
\caption{An example of Amazon Sagemaker team monitoring and supporting SO practitioners' queries (\da{55553190}).}
\label{fig:sagemaker_support}
\end{figure}

Our analysis shows that popular AutoML providers development actively follows SO discussion and provides solutions. For example, the H2O AutoML team officially provides support for practitioners' queries in SO (e.g., \da{51527994}), and thus they provide insights into our current limitations (e.g., \da{52486395}) of the framework and plans for fixing the bugs in future releases (e.g., Fig.~\ref{fig:h2o_support} in \da{51683851}). Similarly, we also find the development team from Amazon Sagemaker also actively participating in the SO discussion (Fig.~\ref{fig:sagemaker_support} in \da{55553190}). In Figure~\ref{fig:top_platform_acc_stat}, we present a bar chart with the total number of questions vs. the percentage of questions with accepted answers for each of the top seven AutoML service providers. We can see that Amazon SageMaker has 2,053 questions, and only 33\% of its questions have accepted answers. From Figure~\ref{fig:top_platform_acc_stat}, we see H2O AI has the highest percentage of questions with accepted answers (42\%), followed by Rapid Miner (40\%), Azure Machine Learning (38\%), Splunk (36\%), Amazon SageMaker (33\%), Google Cloud AutoML (32\%), Amazon Lex (31\%).

\subsection{Traditional ML vs AutoML challenges.} \label{sub_sec:autom_vs_traditionalml}
With with popularity of ML and its widespread adoption, there are several studies on the challenges of machine learning development~\cite{alshangiti2019developing, bangash2019developers}, deep learning development~\cite{han2020programmers}, DL frameworks/libraries~\cite{islam2019developers}. 
However, none of this research addresses the challenges of the low-code machine learning approach, i.e., AutoML. In this section, we compare the conclusions of these past studies on the border ML domain with our findings regarding the AutoML challenges. As AutoML is a subset of the ML domain, we find that many issues are comparable while others are unique. This finding is intriguing in and of itself, given that AutoML was intended to mitigate some of the shortcomings of traditional ML development and make ML adoption more accessible. Our analysis shows that AutoML tools are successful in some circumstances but not all.
For example, among AutoML practitioners, there is substantially less discussion about Model creation and fine-tuning than among all ML practitioners. However, deployment, library, and system configuration dominate both traditional ML and AutoML development. Future research from the SE domain can further ameliorate many of these challenges by providing a guideline on better documentation, API design, and MLOps. We present a detailed comparison of some of the most significant ML development challenges faced by traditional ML versus AutoML practitioners.

\bf{Requirement Analysis.}
This contains a discussion on formulating an ML solution for a specific problem/task. These questions are related to understanding available data and the support for existing ML frameworks/platforms. An empirical study on the machine learning challenges in SO discussion by Alshangiti et al.~\cite{alshangiti2019developing} reports that ML developers have around 2.5\% questions on these challenges in SO. Other studies~\cite{islam2019developers, islam2019developers} also report similar challenges for ML developers.
From our analysis (\sec \ref{rq_result:mllc}), we also find that AutoML practitioners have around 5.7\% queries regarding the supported services from AutoML tools/platforms to meet their business requirements.
Our analysis (Fig.~\ref{fig:bubble_diff_pop_MLLC}) indicates that AutoML practitioners find these to be particularly hard  (e.g., ``Does AzureML RL support PyTorch?'' in \dq{64025308} or ``Is there support for h2o4gpu for deeplearning in python?'' in \dq{67727907}); therefore, AutoML providers should give more support in terms of available resources.

\bf{Data Processing \& Management.}
This challenge relates to data loading, cleaning, splitting, formatting, labeling, handling missing values or imbalanced classes, etc. Previous studies report that ML developers find data pre-processing the second most challenging topic~\cite{alshangiti2019developing} and even some popular ML libraries lack data cleaning features~\cite{islam2019developers}.
Our analysis (\sec\ref{rq_result:topics} and \sec\ref{rq_result:pop_diff}, we find that this is quite a dominant topic for AutoML practitioners (i.e., 27\% questions), but AutoML practitioners find these topics (Fig.~\ref{fig:bubble_diff_pop_per_category}) and MLLC phases (Fig.~\ref{fig:bubble_diff_pop_MLLC}) less challenging. This finding suggests that AutoML service provide better abstraction and APIs for this data pre-processing and management pipeline.

\bf{Model Design.}
This contains some conceptual and implementation-related discussions on ML. This contains a discussion about ML algorithms/models, their internal implementation, parameters, etc. (e.g., Architecture of FCC, CNN, ResNet, etc., learning rate). Similar studies on ML challenges~\cite{islam2019developers} report model creation-related questions as challenging. Questions regarding different ML algorithms are dominant in SO~\cite{bangash2019developers, han2020programmers}, and Neural network architecture is a popular topic among practitioners and gaining popularity~\cite{bangash2019developers}. 
From our analysis, we find that this challenge is prevalent (4.2\% questions) in AutoML practitioners (Fig.~\ref{fig:taxonomy_TM}) and moderately challenging (Table~\ref{tab:topicDifficulty}).

\bf{Training \& Debugging.}
This contains a discussion related to the actual training of the model on a specific train dataset. This contains a discussion about use cases or troubleshooting for a specific library via documentation or prior experience.
Similar studies~\cite{bangash2019developers, islam2019developers, han2020programmers} report that coding error/exception topics  -- type mismatch, shape mismatch -- are most dominant in SO, and resolving these issues requires more implementation knowledge than conceptual knowledge~\cite{alshangiti2019developing}.
Our analysis suggests that AutoML practitioners also find model training and debugging-related questions (i.e., around 10\%) most popular (Fig.~\ref{fig:bubble_diff_pop_topic}). This suggests that both ML and AutoML practitioners require better support for bug detection~\cite{wardat2022deepdiagnosis}.

\bf{Model Evaluation \& Performance Improvement.}
This contains challenges related to evaluating the trained model via different metrics and improving the model's performance via different techniques and multiple iterations. 
Similar studies~\cite{islam2019developers} also report that model-turning and prediction-related queries are less frequent in SO, even though significant ML research focuses on this step. More research is needed on practitioners' challenges.
We find that AutoML practitioners have many queries (i.e., around 10\%) regarding improving the performance of the model with fewer resources (e.g., ``H2O cluster uneven distribution of performance usage'' in \dq{46056853}, \dq{62408707}). AutoML aims to abstract model architecture-related complexities, but better explainability of the model's prediction may mitigate many of AutoML practitioners' concerns (e.g., ``RapidMiner: Explaining decision tree parameters'' in \dq{23362687}).

\bf{Model deployment.}
This contains challenges while deploying the model, scaling up as necessary, and updating the model with a new dataset. 
Chen et al.~\cite{chen2020comprehensive} report that ML developers face several challenges from Model export, environment configuration, model deployment via cloud APIs, and updating model. Guo et al.~\cite{guo2019empirical} evaluated prediction accuracy and performance changes when DL models trained on PC platforms were deployed to mobile devices and browsers and found compatibility and dependability difficulties. Similar studies~\cite{alshangiti2019developing, islam2019developers} on developers' discussion on SO on DL frameworks reported that developers find model deployment challenging~\cite{han2020programmers}.
From our analysis, we also find that (Fig.~\ref{fig:bubble_diff_pop_topic}) AutoML practitioners find Model Load \& Deployment (around 13\% questions) to be one of the most difficult topics (e.g., ``Azure ML Workbench Kubernetes Deployment Failed'' in \dq{46963846}). This is particularly problematic as AutoML is expected to make model deployment and stability-related problems. Future research efforts from SE can help to mitigate these challenges.

\bf{Documentation.}
This challenge is related to incorrect and inadequate documentation, which is also quite prevalent in other software engineering tasks~\cite{aghajani2020software, aghajani2020software}. Similar to empirical studies~\cite{islam2019developers, han2020programmers} on the overall ML domain, we find that API misuse related to inadequate documentation or ML experts~\cite{alshangiti2019developing} is prevalent in all MLLC phases (e.g., In adequate documentation ``API call for deploying and undeploying Google AutoML natural language model - documentation error?'' in \dq{67470364}). This implies that researchers ML/AutoML should collaborate with the SE domain to improve these prevalent limitations~\cite {uddin2019towards}.

\section{Implications}\label{sec:implications}
In this Section, we summarize the findings can guide the following three stakeholders: \begin{inparaenum}[(1)]
        \item AutoML Researchers \& Educators to have a deeper understanding of the practical challenges/limitations of current AutoML tools and prioritize their future research effort,
        \item AutoML platform/tool Providers to improve their services, learning resources, improve support for deployment, and maintenance,
        \item AutoML Practitioners/Developers to gain a useful insights of current advantages and limitations to have a better understanding of the trade-offs between AutoML tools vs Tradition ML tools,
    \end{inparaenum} We discuss the implications in details below.

\textbf{\ul{ AutoML Researchers \& Educators.}}
We find that the AutoML practitioners challenges are slightly different from traditional ML developers (\sec\ref{sub_sec:autom_vs_traditionalml}). Researchers can study how to improve the performance of AutoML core services, i.e., finding best ML algorithm configuration or Neural architecture search. AutoML tools/platforms aims to provide a reasonable good ML model at the cost of huge computational power (i.e., sometimes 100x) as it needs to explore a huge search space for potential solutions~\cite{autoMLcomputation, yang2019evaluation}. Some recent search approach are showing promising results by reducing this cost by around ten fold~\cite{mellor2021neural, elsken2019neural, pham2018efficient, hu2020dsnas}. From our analysis (\sec\ref{rq_diff_pop}), we can see model performance improvement is a popular and difficult topic for AutoML practitioners (Table~\ref{tab:topicPopularity}). Future research on the explanability of the trained model can also provide more confidence to the practitioners~\cite{xu2019explainable, hoffman2018metrics}.Researchers from SE can also help improve documentation, API design, and design intelligent tools to help AutoML practitioners to correctly tag SO questions so they reach the right experts~\cite{bangash2019developers, baltadzhieva2015predicting}.

\bf{\ul{ AutoML Vendors.}}
In this research, we present the AutoML practitioners discussed topics (Fig.\ref{fig:taxonomy_TM}), popularity (Table~\ref{tab:topicPopularity}) and difficulty (Table~\ref{tab:topicDifficulty}) of these topics in details. From Figure~\ref{fig:bubble_diff_pop_topic}, we can see that Model load \& deployment related topic is one of the most popular and difficult topic among the practitioners. We can also see that Model training \& monitoring related question are the most challenging topic and model performance, library/platform management and data transformation related topics are most popular. AutoML vendors can prioritise their efforts based on the popularity and difficulty of these topics  (Fig.~\ref{fig:bubble_diff_pop_per_category}). For example, expect for Model Load \& Deployment (46 hours) and Model Training \& Monitoring (32 hours) topics questions from other 11 topics have median average wait for for accepted answer is less than 20 hours.
We also find that topics under data category are quite dominant and our analysis of AutoML tools and similar studies on other ML libraries~\cite{islam2019developers, alshangiti2019developing} report the limitation of data processing APIs that require special attentions.

From Figure~\ref{fig:bubble_diff_pop_MLLC}, We see that questions asked during model design and training related MLLC phase quite prevalent and popular among the AutoML community. We can also see that AutoML practitioners find requirement analysis phase (e.g., ``How can I get elasticsearch data in h2o?'' in \dq{48397934}) and Model deployment \& monitoring phase quite challenging (e.g., ``How to deploy machine learning model saved as pickle file on AWS SageMaker'' in \dq{68214823}). We can also see popular development teams from popular AutoML providers such as H2O and Microsoft (Fig.~\ref{fig:h2o_support}, Fig.~\ref{fig:sagemaker_support} in \sec\ref{sub_sec:top_automl}) actively follow practitioners discussion on SO and take necessary actions or provide solutions. Other AutoML tool/platforms providers should also provide support to answer their platform feature specific queries. It shows that the smooth adoption of the AutoML tools/platforms depends on the improved and effective documentation, effective community support and tutorials.

\textbf{\ul{AutoML Practitioners/Developers.}}
The high demand and shortage of ML experts is making low-code, i.e., AutoML approach for ML development attractive for organizations. We can also see the the AutoML related discussion in SO a steady upward tread (Fig.~\ref{fig:all_questions_evolution}). AutoML tools/platforms provides a great support for different ML aspects such as data pre-processing and management but still lacks support for various tasks specially tasks related to model deployment and monitoring. Our analysis can provide project managers a comprehensive overview of current status of overall AutoML tools (\sec\ref{sub_sec:top_automl} and \sec\ref{sub_sec:autom_vs_traditionalml}).
Our analysis provides invaluable insight on the current strength/limitations to project managers to adopt AutoML approach and better manage their resources. For example, even with AutoML approach setting up and configuring development environment and library management (e.g., ``Need to install python packages in Azure ML studio'' in \dq{52925424}), deployment trained models via cloud endpoint and updating model is still challenging. Our analysis also shows that average median wait time to get an accepted solution from SO is around 12 hours for data management and model implementation/debugging related tasks as development teams from some of these platforms also provide active support.

\section{Threats to Validity}\label{subsec:validity}

\nd\bf{Internal validity} threats in our study relate to the authors' bias while analyzing the questions. We mitigate the bias in our manual labeling of topics and  AutoML phases by following the annotation guideline. The first author participated in the labeling process, and in case of confusion, the first author consulted with the second author.

\nd\bf{Construct Validity} threats relate to the errors that may occur in our data collection process (e.g., identifying relevant AutoML tags). To mitigate this, first we created our initial list of tags, as stated in Section \ref{sec:methodology}, by analyzing the posts in SO related to the leading AutoML platforms. Then we expanded our tag list using state-of-art approach~\cite{alamin_LCSD_EMSE, alamin2021empirical, iot21, bagherzadeh2019going,abdellatif2020challenges,ahmed2018concurrency, rosen2016mobile}. Another potential threat is the topic modeling technique, where we choose $K$ = 15 as the optimal number of topics for our dataset $B$. This optimal number of topics directly impacts the output of LDA. We experimented with different values of $K$ following related works~\cite{alamin_LCSD_EMSE, alamin2021empirical, abdellatif2020challenges, bagherzadeh2019going}. We used the coherence score and our manual observation to find $K$'s optimal value that gives us the most relevant and generalized AutoML-related topics~\cite{alamin_LCSD_EMSE, alamin2021empirical, iot21, abdellatif2020challenges}. 

\nd\bf{External Validity} threats relate to the generalizability of our research findings. This study is based on data from developers' discussions on SO. However, there are other forums that AutoML developers may use to discuss. We only considered questions and accepted answers in our topic modeling following other related works~\cite{alamin_LCSD_EMSE, alamin2021empirical, iot21, abdellatif2020challenges}. The accepted and best answer (most scored) in SO may be different. The questioner approves the accepted answer, while all the viewers vote for the best answer. It is not easy to detect whether an answer is relevant to the question. Thus We chose the accepted answer in this study because we believe that the questioner is the best judge of whether the answer solves the problem. Even without the unaccepted answers, our dataset contains around 14.3K SO posts (12.5K questions + 3.7K accepted answers). Nevertheless, we believe using SO's data provides us with generalizability because SO is a widely used Q\&A platform for developers from diverse backgrounds. However, we also believe this study can be complemented by including discussions from other forums and surveying and interviewing AutoML practitioners.

\section{Related Work} \label{sec:related_work}

\subsection{Research on AutoML}
There are ongoing efforts to increase the performance of AutoML tools through better NAS approach~\cite{mellor2021neural, elsken2019neural, pham2018efficient, hu2020dsnas}, search for hyperparameters and loss methods~\cite{li2019lfs, li2022volcanoml, li2022volcanoml}, automating machine learning via keeping human in the loop~\cite{lee2020human}.
Truong et al.~\cite{truong2019towards} evaluated few popular AutoML tools on their abilities to automate ML pipeline. They report that at present different tools have different approaches for choosing the best model and optimizing hyperparameters. They report that the performance of these tools are reasonable against custom dataset but they still need improvement in terms of computational power, speed, flexibility etc.
Karmaker et al.~\cite{karmaker2021automl}  provides an overview of AutoML end-to-end pipeline. They highlight which ML pipeline requires future effort for the adoption of AutoML approach.
There are other case studies on the use of AutoML technologies for autonomous vehicles~\cite{li2021automl}, industry application~\cite{li2021automl, li2021automl} anticipate bank collapses, predicting bank failures~\cite{agrapetidou2021automl} where the authors evaluate current strengths and limitations and make suggestions for future improvements.
These studies focus on improving the AutoML paradigm from a theoretical perspective, and they do not address the challenges that AutoML practitioners have expressed in public forums such as SO.

\subsection{Empirical Study on Challenges of ML Applications.}
There has been quite some research efforts~\cite{patel2008examining, sculley_hiddent_debt_2015, SE_DL18, ml_practice_2019} mostly by interviewing ML teams or surveys on the challenges of integrating ML in software engineering,  explanability on the ML models~\cite{roscher2020explainable, jiang2018trust}. Bahrampour et al.~\cite{bahrampour2015comparative} conducted a comparative case of study on five popular deep learning frameworks -- Caffe, Neon, TensorFlow, Theano, and Porch -- on three different aspects such as their usability, execution speed and hardware utilization.

There are also quite some efforts on finding the challenges of Machine learning application by analysing developers discussion on stack overflow.


There are a few studies analyzing developer discussion on Stack Overflow regarding popular ML libraries.
Islam et al.~\cite{islam2019developers} conduct a detailed examination of around 3.2K SO posts related to ten ML libraries and report the urgent need of SE research in this area. They report ML developers need support on finding errors earlier via static analyser, better support on debugging, better API design, better understanding of ML pipeline.
Han et al.~\cite{han2020programmers} an empirical study on developers' discussion three deep learning frameworks -- Tensorflow, Pytorch, and Theano -- on stack overflow and GitHub. They compare and report their challenges on these frameworks and provide useful insights to improve them.
Zhang et al.~\cite{zhang2018empirical} analysed SO and GitHub projects on deep learning applications using TensorFlow and reported the characteristics and root causes of defects.
Cummaudo et al.~\cite{cummaudo2020interpreting} use Stack Overflow to mine developer dissatisfaction with computer vision services, classifying their inquiries against two taxonomies (i.e., documentation-related and general questions). They report that developers have a limited understanding of such systems' underlying technology.
Chen et al.~\cite{chen2020comprehensive} conducts a comprehensive study on the challenges of deploying DL software by mining around 3K SO posts and report that DL deployment is more challenging than other topics in SE such as big data analysis and concurrency. They report a taxonomy of 72 challenges faced by the developers.
These studies do not focus on the challenges of overall machine learning domain but rather on popular ML libraries, ML deployment, or a specific cloud-based ML service. Our study focuses entirely on overall AutoML related discussion on SO; hence, our analysed data, i.e., SO posts, are different.

There are also several empirical research on the challenges of machine learning, particularly deep learning, in the developers' discussion on SO.
Bangash et al.~\cite{bangash2019developers} analyzes around 28K developers' posts on SO and share their challenges. They report practitioners' lack of basic understanding on Machine learning and not enough community feedback.
Humbatova et al.~\cite{humbatova2020taxonomy} manually analysed artefacts from GitHub commits and related SO discussion and report a report a variety of faults of while using DL frameworks and later validate their findings by surveying developers.
Alshangiti et al.~\cite{alshangiti2019developing} conduct a study on ML-related questions on SO and report developers' challenges and report that ML related questions are more difficult than other domains and developers find data pre-processing, model deployment and environment setup related tasks most difficult. They also report although neural networks, and deep learning related frameworks are becoming popular, there is a shortage of experts in SO community.
In contrast, our study focuses on AutoML tools/platforms rather than boarder machine learning domain in general. In this study, we aim to analyse the whole AutoML domains, i.e., AutoML-related discussion on SO rather than a few specific AutoML platforms. Our analyzed SO posts and SO users differ from theirs, and our findings provide insight focus on AutoML's challenges.

\subsection{Research on Topic Modeling \& SO discussion}\label{sec:rel-se-res-topic}
Our reason for employing topic modeling to understand LCSD discussions is rooted in current software engineering study demonstrates that concepts derived from The textual content can be a reasonable approximation of the underlying data~\cite{Chen-SurveyTopicInSE-EMSE2016,Sun-SoftwareMaintenanceHistoryTopic-CIS2015,Sun-ExploreTopicModelSurvey-SNPD2016}.
Topic modelling in SO dataset are used in a wide range of studies to understand software
logging messages~\cite{Li-StudySoftwareLoggingUsingTopic-EMSE2018} and previously for
diverse other tasks, such as concept and locating features~\cite{Cleary-ConceptLocationTopic-EMSE2009,Poshyvanyk-FeatureLocationTopic-TSE2007},
linking traceability (e.g., bug)~\cite{Rao-TraceabilityBugTopic-MSR2011,asuncion2010software},
to understand the evolution of software and source code history
~\cite{Hu-EvolutionDynamicTopic-SANER2015,Thomas-SoftwareEvolutionUsingTopic-SCP2014,Thomas-EvolutionSourceCodeHistoryTopic-MSR2011}, to facilitate categorizing software code search~\cite{Tian-SoftwareCategorizeTopic-MSR2009}, to refactor software code
base~\cite{Bavota-RefactoringTopic-TSE2014}, and for explaining software
defects~\cite{Chen-SoftwareDefectTopic-MSR2012}, and various software maintenance ~\cite{Sun-SoftwareMaintenanceTopic-IST2015,Sun-SoftwareMaintenanceHistoryTopic-CIS2015}. 
The SO posts are used in several studies on various aspects of software development using topic modeling, such as what developers are discussing in general~\cite{barua2014developers} or about a particular aspect, e.g., low-code software developers challenges~\cite{alamin_LCSD_EMSE, alamin2021empirical}, IoT developers discussion~\cite{iot21}, docker development challenges~\cite{haque2020challenges}, concurrency~\cite{ahmed2018concurrency}, big data~\cite{bagherzadeh2019going}, chatbot~\cite{abdellatif2020challenges}, machine learning challenges~\cite{chen2020comprehensive, bangash2019developers, alshangiti2019developing}, challenges on deep learning libraries~\cite{han2020programmers, islam2019developers}.
\section{Conclusions} \label{sec:conclusion}
AutoML is a novel low-code approach for developing ML applications with minimal coding by utilizing higher-level end-to-end APIs. It automates various tasks in the ML pipeline, such as data prepossessing, model selection, hyperparameter tuning, etc. We present an empirical study that provides valuable insights into the types of discussions AutoML developers discuss in Stack Overflow (SO).
We find 13 AutoML topics from our dataset of 14.3K extracted SO posts (question + acc. answers). We extracted these posts based on 41 SO tags belonging to the popular 18  AutoML services. We categorize them into four high-level categories, namely the MLOps category (5 topics, 43.2\% questions) with the highest number of SO questions, followed by Model (4 topics, 27.6\% questions), Data (3 topics, 27\% questions), Documentation (1 topic, 2.2\% questions). Despite extensive support for data management, model design \& deployment, we find that still, these topics are dominant in AutoML practitioners' discussions across different MLLC phases. We find that many novice practitioners have platform feature-related queries without accepted answers. Our analysis suggests that better tutorial-based documentation can help mitigate most of these common issues. MLOps and Documentation topic categories predominate in cloud-based AutoML services. In contrast, the Model topic category and Model Evaluation phase are more predominant in non-cloud-based AutoML services.
We hope these findings will help various AutoML stakeholders (e.g., AutoML/SE researchers, AutoML vendors, and practitioners) to take appropriate actions to mitigate these challenges. The research and developers' popularity on AutoML indicates that this technology is likely widely adopted by various businesses for consumer-facing applications or business operational insight from their dataset. AutoML researchers and service providers should address the prevailing developers' challenges for its fast adoption. Our future work will focus on \begin{inparaenum}[(1)]
\item getting AutoML developers' feedback on our findings by interviews or surveys, and 
\item developing tools to address the issues observed in the existing AutoML's data processing and model designing pipeline.
\end{inparaenum}

\bibliographystyle{ACM-Reference-Format}
\bibliography{bibliography}

\end{document}